\documentclass[journal]{IEEEtran}
\usepackage{amssymb}

\usepackage{caption}
\usepackage{array}
\usepackage{rotating}
\usepackage{subfigure}
\usepackage{multirow}
\usepackage{color}
\usepackage{epsfig}
\usepackage{epstopdf}
\usepackage{url}
\usepackage[linesnumbered, ruled, vlined]{algorithm2e}

\usepackage{amsmath}
\usepackage{amsfonts}
\usepackage{amssymb}

\begin{document}

\title{Interpretable {Feature} Learning in Multivariate Big Data Analysis for Network Monitoring}



 \author{\IEEEauthorblockN{Jos\'e Camacho\IEEEauthorrefmark{1},~\IEEEmembership{{Senior} Member,~IEEE}
		Katarzyna Wasielewska\IEEEauthorrefmark{1},~\IEEEmembership{Senior Member,~IEEE},
		Rasmus Bro\IEEEauthorrefmark{2},
	     David Kotz\IEEEauthorrefmark{3},~\IEEEmembership{Fellow,~IEEE}}
	\IEEEauthorblockA{\IEEEauthorrefmark{1}Research Centre for Information and Communication Technologies (CITIC-UGR), University of Granada, Spain}
	\IEEEauthorblockA{\IEEEauthorrefmark{2}Chemometrics Group, University of Copenhagen, Denmark}
	\IEEEauthorblockA{\IEEEauthorrefmark{3}Department of Computer Science, Dartmouth College, Hanover, United States}
	\thanks{ 
		Corresponding author: J. Camacho (email: josecamacho@ugr.es).}}
	
	\IEEEtitleabstractindextext{%
\begin{abstract}
    There is an increasing interest in the development of new data-driven models useful to assess the performance of communication networks. For many applications, like network monitoring and troubleshooting, a data model is of little use if it cannot be interpreted by a human operator. In this paper, we present an extension of the Multivariate Big Data Analysis (MBDA) methodology, a recently proposed interpretable data analysis tool. In this extension, we propose a solution to the automatic derivation of features, a cornerstone step for the application of MBDA when the amount of data is massive. 
    The resulting network monitoring approach allows us to detect and diagnose disparate network anomalies, with a data-analysis workflow that combines the advantages of interpretable and interactive models with the power of parallel processing. We apply the extended MBDA to two case studies: UGR'16, a benchmark flow-based real-traffic dataset for anomaly detection, and Dartmouth'18, the longest and largest Wi-Fi trace known to date. 
\end{abstract}

}

	\maketitle

	\IEEEdisplaynontitleabstractindextext

	%
	\IEEEpeerreviewmaketitle



\section{Introduction}
\label{sec:introduction}

 \begin{figure*}[t]
 	\centering
	\subfigure[]{\includegraphics[width=0.3\textwidth]{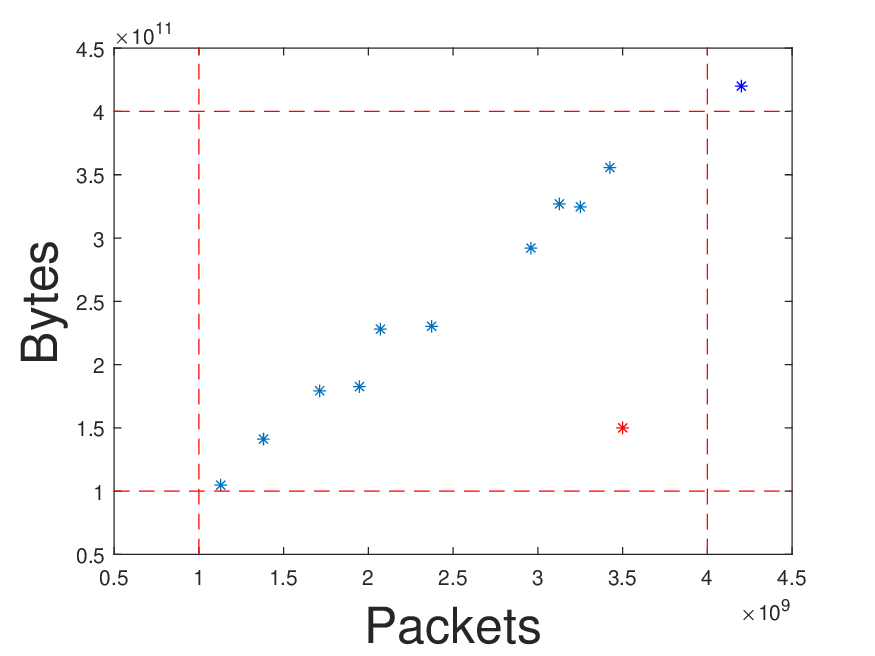}} 
    \subfigure[]{\includegraphics[width=0.3\textwidth]{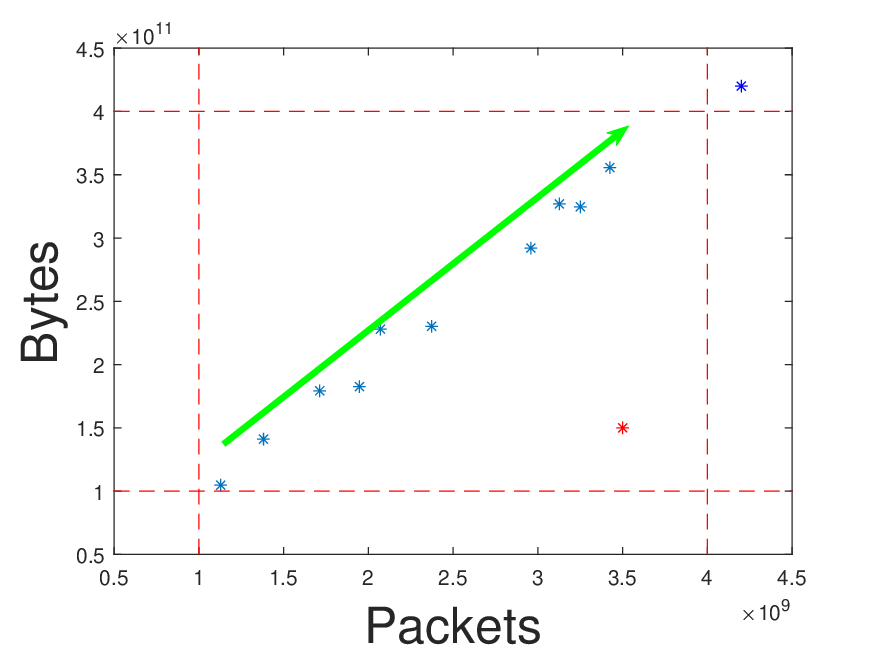}} 
    \subfigure[]{\includegraphics[width=0.3\textwidth]{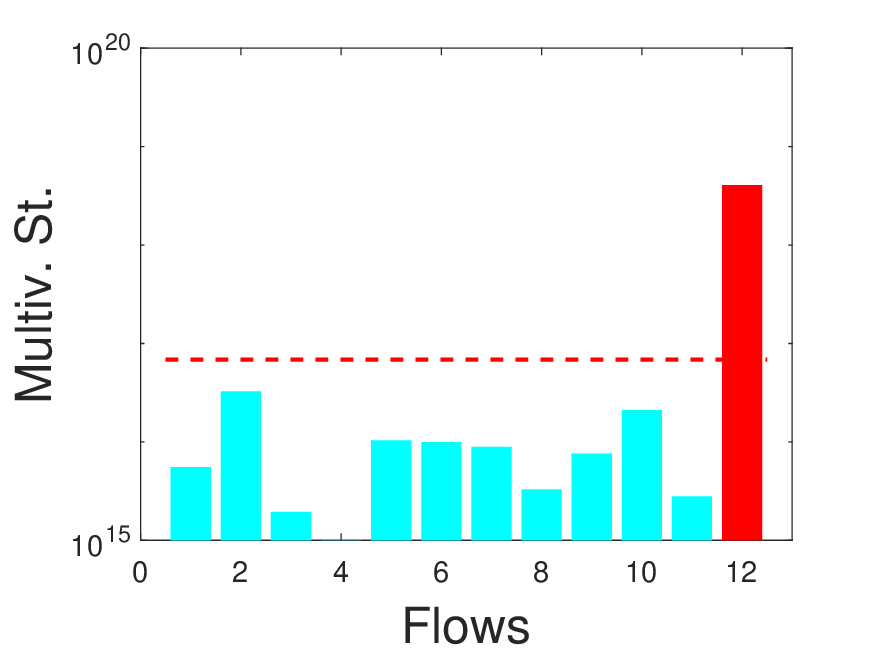}} 
 	\caption{Illustration of a simple multivariate example.}
 	\label{fig:babies}
 \end{figure*}

In the Big Data era, there is an increasing interest in the development of new data analysis methods to improve the performance of communication networks, in tasks like network monitoring, troubleshooting and optimization~\cite{ietf-opsawg-ntf-13}. The current trend in data analysis is towards highly complex black-box methodologies, like deep learning~\cite{pang2021deep}. These methodologies learn models of the data intended to be used automatically, and with little or no human supervision or interaction. Unfortunately, for many network applications, a model of the data is of little use if it cannot be interpreted by a human operator. 

The relevance of interpretable models in several applications has raised a lot of attention in the research community in recent years~\cite{molnar2019interpretable}. There are two basic approaches to the derivation of interpretable models from data. On the one hand, the need for the interpretation of black-box models has given rise to concepts like interpretable or explainable machine learning~\cite{doshi2017towards}, where strategies to explain black-box models or to calibrate more interpretable black-box models are pursued. An alternative approach is to use data analysis methods that are themselves interpretable, rather than black-box~\cite{rudin2019stop}. This paper lies in the second category.

Just like their black-box counterparts, interpretable models can be useful in classification, regression and anomaly detection tasks. However, a major advantage of interpretable models is that they also provide information about \emph{why\/} a model gives a certain output.  There are many situations in which an answer is not of practical use, without knowing the ``why''.  Network monitoring is an example: network operators desire to detect unwanted events during the network operation, but they also need to understand their root causes and troubleshoot them as soon as possible.

Multivariate analysis has been recognized as an outstanding data analysis approach in several domains, including  industrial monitoring~\cite{Ferrer2014}, network security~\cite{Camacho2014}, marketing~\cite{article}, weather modeling~\cite{Jolliffe02}, bioinformatics~\cite{Zou2006}, food research~\cite{Bro98multi-wayanalysis}, and so forth. In this paper, we are interested in a multivariate methodology for data interpretation: matrix factorization with component models. In this methodology, visualization, interpretation and data interaction are the principal tools for an analyst to understand  the problem the data reflects. Two are the main features that make matrix factorization an appealing methodology for the analysis of complex data: i) most matrix factorization models are simple to interpret, because they are based on linear algebra, and ii) they generate factors that simplify the visualization of data. Another advantage is that, even if a model is created to respond to a specific question (e.g., anomaly detection), the interaction of the analyst with the data through the model can bring much more information, like the derivation of new, unexpected findings (e.g., network misconfiguration or sub-optimal functioning). This property is a useful one that black-box models do not normally provide.

Researchers have been quite active in the extension of machine learning methodologies to Big Data. Unfortunately, the extension of multivariate analysis to Big Data while retaining the capabilities of visualization, interpretation and data interaction has received little attention. In this context, the Multivariate Big Data Analysis (MBDA) tool~\cite{mbda} is a recent multivariate anomaly detection and data analysis approach suitable for Big Data. It is based on three modules: the {upstream} module, which transforms a Big Data stream into a small feature data; the analysis module, where the analyst can interact with the featured data to analyze and interpret anomalies; and the {downstream} module, useful to map anomalies to the original logs in the Big Data stream, so that operators can derive full understanding of their root causes.  
MBDA works as a magnifying glass into massive amounts of data,  with a configurable trade-off between the level of detail for data visualization and the capability for data compression. The key to  this trade-off is the {upstream} module, where we set the features and the time resolution for the subsequent analysis. In the original MBDA proposal~\cite{mbda}, the features were manually defined, which is a sub-optimal solution and complicates its application to truly massive volumes of data. 

In this paper, we define an automatic feature learning procedure { that can be specially useful in combination with} MBDA. 
This enhancement improves the performance in relatively massive datasets, and is {fundamental in datasets massive and/or complex enough so} that manual features cannot be properly defined due to inherent limitations in the screening of raw data. To illustrate the resulting methodology, we present two case studies: i) a capture from a real network of a tier 3 Internet Server Provider (ISP)~\cite{macia2018ugr}, and ii) a campus-wide Wi-Fi network~\cite{wifi}. {We include both data sets to provide a general evaluation of our feature learning approach. The first data set includes labeled test data with a number of attacks that we leverage to assess the convenience of the automatically generated features in anomaly detection using standard performance measures. The second data set has never been visualized due to its challenging nature except by univariate time series~\cite{wifi}, and we use it to showcase how our approach can help improve the understanding of the data.}

Our contributions in this paper are as follows.

\begin{itemize}
		\item We contribute an automatic feature-learning procedure, consistent with the MBDA methodology.
		
		\item We integrate this procedure into a Python tool and make it available for the community. This Python tool allows the parallelization of the computation in high-performance processing centers.
		
		\item We showcase the extended MBDA approach with feature learning in two real case studies, one from structured netflow data and one from unstructured SNMP data, highlighting what the method can provide to network operators and presenting the workflow in detail.

\end{itemize}

The rest of the paper is organized as follows. {Section~\ref{sec:related} contextualizes our research in the literature.} Section~\ref{sec:interpretable} discusses the interpretable and interactive characteristics in multivariate analysis. 
Section~\ref{sec:mbda} presents the MBDA methodology. Sections~\ref{sec:fl} describes the interpretable learning approach proposed in this paper. Section~\ref{sec:wifi} introduces the materials and methods of the experimental study.
 Sections~\ref{sec:ugr} and \ref{sec:action} walk through the case studies. 
Section~\ref{sec:conclusion} provides final conclusions.

{
\section{Related Works}
\label{sec:related}

Due to the increasing complexity of networks and the growing trend in network traffic, network monitoring has become increasingly challenging~\cite{8789667}.  The literature is rich in data sources and methods for monitoring~\cite{9377998,9108976,9203310}. Fuentes-García \emph{et al.}~\cite{9381201} discuss the framework of data integrators in network security monitoring, with the notable example of Security Information and Event Management (SIEM) Systems. In such systems, sensors of different nature (sensors of traffic, logs and system state, and security sensors) deployed throughout the network send data to a centralized integrator that unifies it for event correlation and alarm triaging. The Network Telemetry Framework (NTF)~\cite{rfc9232} is a standardization effort to enable real-time and fine-grained network monitoring for autonomic networks~\cite{chaparadza2022autonomic,behringer2015rfc}. An example of application of this concept is presented in~\cite{Fezeu2020}. In~\cite{Sivanathan2020} the authors present the results of experiments on programmable telemetry. The notion of network observability has been behind network monitoring and management practices from the early times. In the era of Big Data, observability is not only a matter of devising the best data measurement techniques, but also of properly engineering good practices for data visualization, exploration, and understanding. The methodology proposed in this paper responds to this need by allowing the automatic identification of relevant features that can be integrated within MBDA to provide full network observability through massive data analysis, following a systematic and data-agnostic workflow for exploration and diagnosis. 

Feature learning is an important topic in machine learning research~\cite{Bengio2013}. N-gram models represent the traditional feature learning approach in natural language processing (NLP), probably one of the most popular machine learning applications. N-grams are counts extracted from a corpus of text for series of $N$ adjacent letters (or other language units), where $N$ is user-defined. Thus, letter-based N-grams treat words as atomic units. \emph{Word2vec}~\cite{mikolov2013efficient} is an extremely popular feature learning approach for NLP tasks that extends N-gram models by defining word embeddings, that is, transformations from words to feature vectors that capture the relationship among words in the sentences of a corpus of text. Word2vec looks for a compromise between computational complexity and modelling performance in comparison to (deep) neural networks, where the word embedding is part of the machine learning model itself~\cite{Cheng2022,Shone2018}
.  By separating the feature learning from the model training, word2vec is capable of extracting general features that can be used in different NLP applications and models. There are several word embedding strategies that follow a similar approach~\cite{Pennington2014,Bojanowski2017,WangJin2020,Ryciak2022}. The proposal of this paper is similar in philosophy to the N-gram approach, and relays in multivariate analysis to model the relationship among words. This allows for a benefit in computing power, and derives features that are even more application agnostic than those from word2vec. Most importantly, our features are as interpretable as N-grams, which is a major goal of this paper. Unlike N-grams, our approach neither intends to model all possible combinations of $N$ letters, nor assumes a specific number of letters for each feature. This is achieved through the definition of the concepts of variable and feature, as we discuss below. This approach
makes feature vectors more flexible and parsimonious, but yet very powerful for data modelling thanks to the concept of default features, which allows us to model residual information of variables that is not included in any learned feature.

A special type of feature learning is the one specifically defined for networks of interconnected nodes, like social networks, biological networks, and communication networks. This methodology, often referred to as network feature learning, assumes that the data represents the information shared among network nodes, which is appropriate for traffic data and many forms of security data, but not for application and system logs. A popular approach for network feature learning, inspired in the {word2vec} algorithm, is \emph{node2vec}~\cite{grover2016node2vec}. Network feature learning with node2vec and related node embeddings~\cite{Schlotterer2019,Salamat2020} can be useful in tasks such as graph, node or link classification, prediction, and anomaly detection. Generally speaking, our approach has a wider  applicability than network feature learning, since it can be applied to any form of monitoring data. 

Interpretablity is a major need in network operations, especially when there is a need for accountability of management decisions~\cite{kim2020transparency}. For proper accountability, a full understanding of the behavior of the network is required. As already discussed, there are two basic approaches to the derivation of interpretable models: interpretable or explainable machine learning~\cite{doshi2017towards,Du2019,Moradi2021,Tahmassebi2020,Li2020} and the use of data analysis methods that are themselves interpretable~\cite{rudin2019stop, MSNM2016, Ferrer2014}. Moreover, it is worth noting that the self-explaining AI approach has also been proposed as a solution to interpretability~\cite{Elton2020}. An example of explainable feature learning in the network context can be found in~\cite{kang2022providing}, where node2vec is combined with reinforcement learning. The basic idea of such approaches is to maintain the performance of the learning methodology, but provide means of explanation of model outcomes. In this paper, we are more interested in the alternative research avenue, where we retain the interpretability of white-box models but look for an improvement of performance through learning. 

There are other interpretable methods different to multivariate analysis through matrix factorization; see for instance~\cite{molnar2019interpretable}. Useful interpretations can be derived from statistical methods (e.g., wavelet analysis, covariance matrix analysis), clustering methods (KNN, k-means), decision trees, decision rules and bayesian methods, among others.  While the use of PCA for dimensionality reduction is well-known, the interpretability capabilities of PCA~\cite{Jolliffe2016} have not been exploited in the network domain. In this paper, we develop our feature learning approach so that interpretability is maintained at the expense of generating feature vectors of large dimensionality, which can be easily accommodated by multivariate analysis. This makes our learning features especially well suited for PCA but also other multivariate approaches, like Partial Least Squares (PLS) regression~\cite{W&S&E01}, ANOVA Simultaneous Component Analysis (ASCA)~\cite{smilde2005anova}, Parallel Factor Analysis (PARAFAC)~\cite{harshman1994parafac} or Sparse methods~\cite{Jolliffe2003, Zou2006}, to mention just a few.

}

\section{Interpretability and Interaction in Multivariate Analysis}
\label{sec:interpretable}

This section is intended to motivate why and how multivariate analysis can be useful in the analysis of Big Data streams. The core of the original MBDA~\cite{mbda} is the Multivariate Statistical Network Monitoring (MSNM)~\cite{MSNM2016} approach, which is originally based on Principal Component Analysis (PCA)~\cite{Jolliffe02,Jackson03}. PCA is the most extended, most simple and most general matrix factorization. Here, simple and general are interesting features, since PCA will be easy to interpret and applicable to almost any data set. MSNM is a PCA-based approach for anomaly detection grounded on the theory of statistical control developed in the process industry by the end of the previous century~\cite{Kresta1991,Nomikos1994,Ferrer2007}. Interpretability and data interaction constitute the foundation of this methodology. 

{
\subsection{The benefit of going multivariate}

{ Most network datasets are originally multivariate, in the sense that they are formed by a number of distinct variables (like number of flows, number of packets, delay measurements, etc.) or by the same variable captured in different locations. Even if a dataset is originally a univariate time series, like a traffic capture, it can be transformed into a multivariate feature dataset by properly computing a number of features, as it is done in the Netflow protocol.} 

In Figure \ref{fig:babies}, we illustrate with a very simplistic example the advantage of looking at data from a multivariate perspective. Let us take an hypothetical example of traffic in a single network link, collected using flow-level statistics. The leftmost plot in the figure shows a scatter plot of eleven flows in terms of the number of packets and bytes. These numbers are most often correlated, so that the number of bytes tend to grow with the number of packets, while the exact correlation will differ from link to link and in time. In this hypothetical sample of flows, most of them follow the same trend, except for the one highlighted in red color, which clearly includes a lower number of bytes than the one expected from the number of packets. Note this specific behavior does not need to be relevant from a management standpoint, but it showcases a pattern that can only be observed from a multivariate perspective. To see this, we included hypothetical upper and lower bounds on the numbers of bytes and packets, that can be observed as dashed horizontal and vertical red lines. For the red flow, both the number of packets and of bytes are within the normal range, so looking at each variable separately will not allow to identify that the red flow is singular. In general, any univariate time series model or statistical chart does not allow to identify multivariate patterns. 

A multivariate model of the sort we consider in this paper is represented by the green arrow in the second plot in Figure \ref{fig:babies}. The model is automatically trained to approximate the blue points as much as possible, and so the latent multivariate structure represented by them. The model of the example is very simple, but in a real situation very complicated solutions can be automatically implemented.  

From the multivariate model we can compute multivariate statistics and control charts capable to automatically adapt to the latent structure in the data, and to find complex anomalies that do not follow such structure. One simplistic approach to build a multivariate statistic is to measure the distance of each point to the model (arrow). In our example, this would give us the rightmost plot in Figure \ref{fig:babies}, in which we can add a control limit to identify anomalous flows. In this plot, the distance to the model of a flow is represented by the height of the corresponding bar, and the red flow (bar) can be clearly identified as anomalous. 

This multivariate approach scales very nicely with the number of original variables, and allow us to automatically train multivariate models that capture the latent multivariate trends in the data, and to find objects that do not follow those. These patterns are abundant in traffic, and generally in network data, and they cannot be spotted by looking at univariate series and charts, like traditional boxplots.     

}

\subsection{PCA Matrix Factorization for Interpretation}

Let us take a data matrix $\mathbf{X}$ with $N$ rows and $M$ columns.
The rows represent the observations (a.k.a individuals, objects or items). Generally speaking, observations are the elements one would like to compare, in order to understand their differences and commonalities. The columns of the data matrix represent the variables (a.k.a. features) that are measured per observation. 

PCA transforms matrix $\mathbf{X}$ into a number $A << M$ of uncorrelated features: the so-called principal
components (PCs). The PCs are ordered by captured variance.
PCA follows the expression:

\begin{equation} \label{eq:PCAm}
\mathbf{X} = \mathbf{T}_{A} \cdot \mathbf{P}_{A}^{t} +
\mathbf{E}_A,
\end{equation}

\noindent where $\mathbf{T}_{A}$ is the $N \times A$ \emph{scores} matrix containing the projection of the observations in the PCs sub-space, $\mathbf{P}_{A}$ is the $M \times A$ \emph{loadings} matrix containing the linear combination of the variables represented in each of the PCs, and $\mathbf{E}_A$ is the $N \times M$ matrix of \emph{residuals}.

We call model (\ref{eq:PCAm}) a matrix factorization, since the information in $\mathbf{X}$ is factorized into the scores in $\mathbf{T}_{A}$,  the loadings in $\mathbf{P}_{A}$ and the residuals $\mathbf{E}_A$. While in the Machine Learning discipline, PCA has been traditionally regarded as a simple pre-processing mechanism to handle high-dimensional data, the matrix factorization in Eq. (\ref{eq:PCAm}) is especially useful for the visualization of complex data. Thus, we can explore the distribution of the observations (rows) and of the variables (columns) of $\mathbf{X}$ in separate plots of $\mathbf{T}_{A}$ and $\mathbf{P}_{A}$, respectively. The latter are of much lower dimension than $\mathbf{X}$, and hence easier to visualize, while they retain most of the information in the data.

The plots of  $\mathbf{T}_{A}$ are called score plots, while the plots of  $\mathbf{P}_{A}$ are called loading plots. Clusters, trends or outliers can be identified in the plots. We can also combine scores and loadings in a single plot, commonly called a \emph{biplot}~\cite{Gabriel71}. Well-designed biplots allow us to establish the interaction between observations and variables. If one observation is located  close to a variable in the biplot, we expect this observation to have a high value (load) of that variable. This property is useful to draw connections between the patterns of observations and variables: e.g., to identify which variables make an outlier different from the rest of observations. { The interested reader can find an example of exploratory analysis with PCA in the Supplementary Materials.}

The matrix factorization in PCA can be extremely useful to understand data sets of high dimensionality, with up to thousands of variables or even more. Data interaction is also central in matrix factorization, due to its reduced computational burden: we can create a specific model to study in detail any pattern we find, or we can discard the data in a pattern in order to find new and more subtle patterns. 

\subsection{MSNM for Interpretable Anomaly Detection}


MSNM is an extension of the Multivariate Statistical Process Control developed in the past century, and originally inspired by the pioneering work in industrial quality control by Walter Andrew Shewhart~\cite{MSNM2016}. MSNM is based on the PCA analysis of network data (traffic, logs, etc.), previously codified as interpretable counters. As part of statistical theory, interpretation has been a major cornerstone of MSNM. 



MSNM handles the high-dimensional network data with PCA. 
From the scores and residuals in PCA, the data is further compressed in a pair of statistics, the {D-statistic} (D-st) and {Q-statistic} (Q-st), that represent the normality level of an observation in the model and residual sub-spaces of PCA, respectively. Upper control limits (UCLs, thresholds) are defined for each statistic to facilitate the detection of anomalies~\cite{MSNM2016}. UCLs leave below-normal observations with a certain confidence level, e.g., $99\%$. 
An anomaly is detected if either its D-statistic or its Q-statistic exceed the corresponding control limit.  {An illustration of a multivariate chart of an statistic like the {D-statistic} and the {Q-statistic} is in the rightmost plot of Figure \ref{fig:babies}}.

The {D-statistic} and the {Q-statistic} for observation $n$ are computed with the following equations:
	\begin{equation}
	D_{n} =  \mathbf{t}_{n} \cdot (\Sigma_T)^{-1} \cdot \mathbf{t}^t_{n} 
	\end{equation}
	\begin{equation} \label{qst}
	Q_{n} = \mathbf{e}_{n} \cdot  \mathbf{e}^t_{n}
	\end{equation}
\noindent where $\mathbf{t}_{n}$ is a $1 \times A$ vector with the scores for observation $n$, $\mathbf{e}_{n}$ is a $1 \times M$ vector with the residuals, and $\Sigma_T$ represents the covariance matrix of the scores. In order to detect anomalies, the number of {PCs to use has to be determined. There are many} methods to aid in that decision~\cite{Jackson03,Saccenti201599}.

Once an anomaly is detected, its interpretation is necessary for root cause analysis. Interpretation of anomalies in MSNM can be done following different diagnostic approaches~\cite{Alcala2009,FUENTESGARCIA2018194}, but all of them amount to identifying a subset of variables associated with the specific anomaly. Generally speaking, diagnostic plots are plots where the contribution of the set of variables to a single statistic (D-st or Q-st) can be inspected.



\section{Multivariate Big Data Analysis}
\label{sec:mbda}

\begin{figure*}[htbp]
	\centering
	\includegraphics[width=0.8\textwidth]{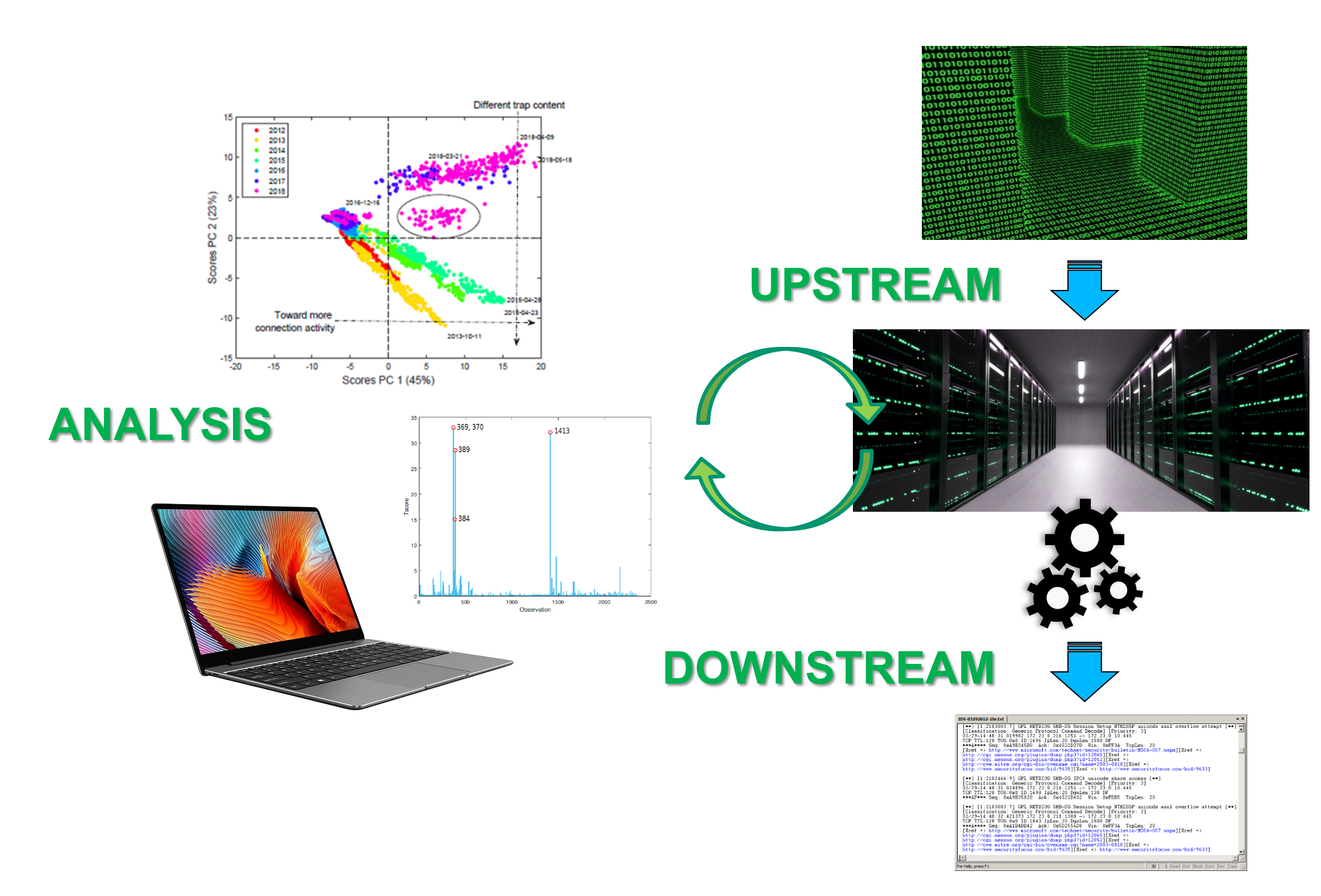}
	\caption{ Multivariate Big Data Analysis diagram: upstream phase, analysis phase and downstream phase. The first and last phases are performed in a cluster of computers or powerful server. The analysis can be performed on a regular computer. Comic image from www.slon.pics~\cite{SLON} in Freepik.}
	\label{fig:trap}
\end{figure*}

The MBDA approach is depicted in Figure \ref{fig:trap}. It consists of three stages: {upstream}, analysis and {downstream}:

\begin{itemize}
	\item[1)] In the {upstream} stage, we transform the Big Data input stream, coming from structured and/or unstructured sources, into time-resolved counters. The input stream is the data collected from the network (e.g., through a Security Information and Event Management system): typically a massive amount of logs and messages stored in a collector, potentially including different sources like network traffic, routing logs, SNMP, etc.~\cite{9381201}. We transform this data into a compressed form we refer to as the feature data. If several sources of data are considered, the features of the different sources of data are combined into a single feature data stream~\cite{MSNM2016}. 
	
	\item[2)] In the analysis module, we visualize the feature data to identify anomalies in time using PCA and MSNM. For each anomaly, we find the associated features with a diagnosis plot, which provides a fast first hint to understand its root causes. The output of this second stage is a list of anomalies identified in time and the associated features. 
	
	\item[3)] De-parsing: Using both detection and diagnosis information, we identify the original raw data records out of the massive input data that are related to the anomalies. This list of records allows a more detailed diagnosis, providing information about specific IPs, ports, etc. involved in the anomaly. The original MBDA paper~\cite{mbda} reports an accuracy above $0.99$ in presenting anomalous records, drastically reducing the amount of information to inspect by the human operator. 
	
\end{itemize}

MBDA makes use of two open software packages available on Github: the MEDA Toolbox~\cite{Camacho201549,MEDARep} and the FCParser~\cite{FCPRep}. The FCParser is a Python tool for the parsing of both structured and unstructured logs. The MEDA Toolbox is a Matlab/Octave toolset for multivariate analysis and data visualization. The FCParser is used in the {upstream} and {downstream} modules, potentially on top of a computer cluster with enough computing power to handle the Big Data stream. The MEDA Toolbox is used in the analysis module in a regular computer, simplifying the interactive analysis by the human operator. 

Basically, the {upstream} module transforms a Big Data stream into a manageable feature data set, that can be analyzed interactively in a traditional computer with multivariate analysis tools. Any interesting pattern found during the analysis can then be contrasted with the raw data thanks to the {downstream} module. Following this approach, we retain the interpretability and interactive nature of multivariate methods for the analysis of Big Data streams. These characteristics constitute a major advantage of MBDA over other Big Data methodologies, in particular black-box models.

\subsection{Feature-as-a-counter parsing}

In the {upstream} stage, network logs are transformed into feature data. MBDA makes use of the feature-as-a-counter (FaaC) approach~\cite{Camacho2014}, described below.

In FaaC, each feature contains the number of times a given event takes place during a pre-defined time interval. Examples of suitable features are the counts of a given word in a log or the number of traffic flows with given destination port in a \textit{Netflow} file. This flexible feature definition makes it possible to integrate, in a suitable way, most sources of information{, and it is similar to state-of-the-art approaches in Natural Language Processing, where n-grams and words are regularly used. } 

To implement the FaaC, the FCParser defines \textit{variables} and \textit{features}. Variables represent general entities in the raw data. In the previous two examples, the variables would be \textit{word} and \textit{destination port}. The features are defined for a specific value or regular expression of a variable. Examples of features would be \textit{word=`food'} and \textit{destination port=`80'}. This representation in variables and features has the relevant advantage that allows for the definition of \textit{default} features, e.g., \textit{word=$<$ANY\_OTHER$>$}, useful to count the instances of a variable that have not been considered in another feature. 

Variables and features are defined using regular expressions in configuration files, where we also set the time resolution of the parsing. 
Each configuration file typically contains several variables and several features per variable. The FCParser applies this configuration to the data to compute a feature vector for each interval of time present in the original data. This operation is done using a multi-threading configuration to speed-up computation.  By selecting the time resolution and the features, we define the trade-off between level of detail and compression. Defining more features and/or using a lower time resolution result in more detail, while defining fewer features and/or using a higher time resolution lead to more compression. 

{
A couple of examples of the FaaC approach can be found in Supplementary Materials. A more detailed example can be found in the FCParser manual~\cite{FCPRep}.} 


\section{Feature Learning in the {upstream} Stage}
\label{sec:fl}

MBDA relies on the definition of the features in the configuration files of the FCParser. To write such configuration files, the analyst needs to get familiarized with the data. 
Unfortunately, in a practical Big Data problem like the ones under analysis, the data capture is simply too massive {and complex, with varying information along time,} for direct inspection.  If we want to obtain a good description of the content, we may apply an automatic feature-derivation technique. The definition of this technique is not straightforward, since it needs to be consistent with the subsequent multivariate analysis, so that we maximize compression while retaining the interpretability required for anomaly detection and root cause analysis. There are two basic properties we would like to have in the learning procedure in consistency with PCA and MSNM: i) the main sources of variance (patterns of change within the data) need to be captured, and ii) uncommon characteristics with low variance should also be modeled somehow, in a summary of residual information. {Furthermore, since we are using multivariate analysis, we can define relatively large numbers of features. Actually, the more features we define, the more descriptive the root cause analysis.} 


{The main contribution of this paper is} a learning algorithm to automatically identify a list of common FaaC features in a Big Data set. {We also included that algorithm} in the FCParser repository at Github with the name \texttt{fclearner.py}. 
The learning algorithm is depicted in Algorithm \ref{alg}. It takes as input a data set and a configuration file with the regular expressions of the {\it variables}. {The learning algorithm starts from this definition of the variables, learns features that show a minimum prevalence in the data, and automatically adds them to the configuration file. This configuration file is then used in the upstream phase to transform the raw data into counters using the FaaC approach. The prevalence of a feature is defined as the portion of {log entries} where the feature appears in the raw data. Thus, if for instance the input configuration file includes the variable  \textit{destination port}, the algorithm will make a list of all possible destination ports in the data, and retain (learn) only those that are present in more than a predefined percentage of log entries.

In our algorithm, we define two prevalence user-defined thresholds. A feature needs to satisfy both thresholds to be included in the configuration file. The thresholds assess the local and global prevalence, $T_l$ and $T_g$, respectively. $T_l$ controls the minimum prevalence in a single time interval, hence the name {\it local}. Any feature that shows a prevalence of at least $T_l$ in a single time interval satisfies this threshold, regardless this feature is not found in any other interval. $T_g$ controls the minimum prevalence in the whole data capture. 

Take the hypothetical example in Table \ref{tab:exLearner}, where the flow-level traffic of a link is broken down in some specific destination ports. Let us assume we would like to automatically learn features in this data for the input variable ``destination\_port'', that we define $T_g=0.05$ and $T_l = 0.2$, and that no other destination port that is not listed in the table is relevant. The most prevalent port is clearly HTTP. Its global prevalence is $408/976 = 0.418$, and its maximum local prevalence is $44/82 = 0.5366$, yielded in the last interval. Since those two values are above the corresponding thresholds, HTTP (destination port 80) is included as a feature in the output configuration file. SMTP  attains a global prevalence of $99/976 = 0.1014$ and a maximum local prevalence of $11/82 = 0.1341$, also for the last time interval. In this case, the global prevalence satisfies $T_g$, but the maximum local prevalence is below $T_l$, reason why SMTP  (destination port 53) is discarded and not included into the set of learned features. Finally, SSH attains a global prevalence of $66/976 = 0.0676$ (above $T_g$) and a maximum local prevalence of $20/81 = 0.2469$ (above $T_l$) found in time interval 8. Even if this feature has less global prevalence than SMTP, it is indeed more interesting given the abnormally high value in a single interval. Since SSH (destination port 22) satisfies both thresholds, the learning algorithm includes it in the final set of features. Note that an interesting property of this learning algorithm, unlike alternative feature engineering approaches like Word2Vec~\cite{mikolov2013efficient,mikolov2013distributed}, is that all learned features can be easily interpreted as a single realization of a variable of the list we provide as input, e.g. for variable ``destination\_port'', a potential feature would be ``destination\_port\_80''. This interpretability is a cornerstone for the combination of the proposal with MBDA.

\begin{table}[t]
	\caption{Hypothetical example of the learning algorithm.}
	\label{tab:exLearner}
	\centerline{
		\small{
			\begin{tabular}{l c c c c}
				\hline \\[-1.5ex]
				{\bf Time interval} & {\bf Total Flows} & {\bf HTTP (80)} & {\bf SMTP (53)} & {\bf SSH (22)} \\[0.5ex]
				\hline \\[-1.5ex]
				1 & 89  &  40  &   9 &    6\\
				2 & 101 &   39 &   10 &    5\\
				3 & 107 &   41 &    9  &   5\\
				4 & 126 &   41 &    9   &  6\\
				5 & 93  &  37  &  10   &  5\\
				6 & 102 &   40 &   12  &   5\\
				7 & 99  &  39  &   9   &  5\\
				8 & 81  &  43  &  10   & 20\\
				9 & 96  &  44  &  10   &  5\\
				10 & 82  &  44  &  11   &  4\\
				\hline \\[-1.5ex]
				{\bf  Total } &  {\bf  976 }&  {\bf408 }&   {\bf99 }&   {\bf66}\\[0.5ex]
				\hline
			\end{tabular}
	}}
	
\end{table}   
}

Both local and global thresholds are complementary. The local threshold $T_l$ needs to be satisfied at least in one time interval. The global $T_g$ needs to be satisfied in the complete data set. Satisfying both thresholds implies that any feature learned has to show a prevalence above $T_l$ in at least one interval and a global prevalence above $T_g$. That way, we learn those features that may be related to anomalous patterns in a handful of intervals, but with enough relevance to be considered a main source of change, meeting our first aforementioned requirement (i)). Those non-learned counters will still be integrated into the default features, so that we still have (arguably limited) observability of low variance patterns, meeting our second requirement (ii).     

The learning algorithm works as follows. For each variable in the configuration file, the algorithm extracts its different features ($F_j^1$ to $F_j^f$) and the number of records in which they are found (counts $\#F_j^1$ to $\#F_j^f$) and store them in $\mathbf{P}_j$. The features which prevalence is above $T_l$ in at least one interval are included in list $\mathbf{F}_j$. At the last part, features with a global prevalence below $T_g$ or that are not in the list are discarded and their prevalence accumulated in the corresponding default feature ($F_j^{def}$) of the variable. Finally, the learning algorithm outputs all the features that satisfy both thresholds and the default features. The \texttt{fclearner.py} tool that implements this algorithm automatically transforms this output in a FCParser configuration file. In turn, this file can be used in the {upstream} stage of MBDA. 

\begin{algorithm}
	\caption{Pseudocode for the learning algorithm.} \label{alg}
			\begin{tabbing}
			IN\=PUT:\\
			\>  $\mathbf{V} \leftarrow \{ V_1, ..., V_v \}$: Regular expressions of variables \+\\
			$\mathbf{D}  \leftarrow \{ D_1, ..., D_d \}$: Data files of disjoint time intervals\\
			$T_l$: Local threshold \\
			$T_g$: Global threshold \-\\ \\
			Ini\=tialization:\\
				\> Set $C$ = 0: global counter of entries  \+ \\
				 Fo\=r each variable $j$  \\
					\> Set $\mathbf{P}_j = \emptyset$ \+  : pairs of features and counts \\ 
					Set $\mathbf{F}_j = \emptyset$ : list of features above threshold \\
					Set $\#F_j^{def}$ = 0: count for default feature \- \- \\  \\
					
			Al\=gorithm:\\ \\
				\> Fo\=r each data file $D_i$ \+ : parallelize at this point\\
					\> Fo\=r each time interval $D_i(t)$ in $D_i$ \+  \\
						\> $C_i(t) \leftarrow  count\_entries(D_i(t))$ \+\\
						$C \leftarrow  C + C_i(t)$ \\
						Fo\=r each variable $j$  \\
							\> $ \{ (F_{j}^1,\#F_{j}^1) ..., (F_{j}^F,\#F_{j}^F) \}  \leftarrow  Match(V_j, D_i(t)) $\+\\
							$\mathbf{P}_j  \leftarrow  combine(\mathbf{P}_j,  \{ (F_j^1,\#F_j^1) ..., (F_j^f, \#F_j^F) \}$ )\\
							 Fo\=r $F_{j}^f$ in $\{ F_{j}^1 ..., F_{j}^F \}$ \\
							      \> If \= $(\#F_{j}^f / C_i(t)) > T_l$ \+ \\
							  		    \> $\mathbf{F}_j  \leftarrow  \mathbf{F}_j \cup F_j^f$ \- \- \- \- \\
							  		    \\
							  		    
				Fo\=r each variable $j$  \\
			       \> Fo\=r \= $F_j^f$ in $\mathbf{P}_j$ \+ \\
			       		\> If \= $(\#F_j^f/C) < T_g$ OR $F_j^f \not \in \mathbf{F}_j$ \+ \\
			              	\> $\#F_j^{def} \leftarrow \#F_j^{def} + \#F_j^i$ \+\\
			              	   $\mathbf{P}_j \leftarrow discard(\mathbf{P}_j,  (F_j^i,\#F_j^i)$) \- \- \- \\ \\

			OUT\=PUT: $\mathbf{P}_j$ for all variables
			
		\end{tabbing}
\end{algorithm}

\section{Materials \& Methods}
\label{sec:wifi}

Below we describe the experimental case studies and the computational architecture used.

\subsection{The UGR'16 Case Study}

{The UGR'16 dataset~\cite{macia2018ugr}\footnote{Dataset available online at \url{https://nesg.ugr.es/nesg-UGR'16/}} was captured from a real network of a tier 3 Internet Server Provider (ISP). The data collection was carried out with Netflow between March and June of 2016 under Normal Operation Conditions (NOCs), meaning that the network was used normally by the ISP clients. This allowed to model and study the normal behavior of the network, and to unveil certain anomalies such as SPAM campaigns. The flows of the dataset were labelled indicating if they were ``background" (regarded as legitimate flows), or ``anomalies" (identified as non-legitimate flows). In addition, another capture was made between July and August of 2016, including some controlled attacks that were launched to obtain a test dataset for validation of anomaly detection algorithms. To do this, twenty five virtual machines were deployed within one of the ISP sub-networks. Five of these machines attacked the other twenty. The type of attacks were \textit{Denial of Service} (DOS), \textit{port scanning} in two modalities: either from one attacking machine to one victim machine (SCAN11) or from four attacking machines to four victim machines (SCAN12), and \textit{botnet traffic} (NERISBOTNET). In this second capture, the ``anomalies" correspond to the synthetic attacks, and the rest of flows are categorized as ``background".
	
As of today, the UGR'16 has been referenced in more than 180 research papers (according to Google Scholar) and it can be considered a benchmark in the research of anomaly detection in real traffic data for cybersecurity. The general characteristics of the dataset are provided in Table \ref{tab:setsFeatures}. To obtain more details on the data, the reader is referred to the original paper~\cite{macia2018ugr}.
	
Using the FaaC approach, we performed anomaly detection at 1 minute intervals rather than at flow level. A total of 134 features were extracted per interval. The process of feature extraction was based on two steps: i) binary files were transformed to flow-level csv files with the nfdump tool, and ii) csv files were transformed to feature vectors with the FCParser.}

\begin{table}[t]
	\caption{Characteristics of the calibration and the test sets in UGR'16.}
	\label{tab:setsFeatures}
	\centerline{
		\small{
			\begin{tabular}{l c c}
				\hline \\[-1.5ex]
				{\bf Feature} & {\bf Training} & {\bf Test} \\[0.5ex]
				\hline \\[-1.5ex]
				Capture start & 10:47h 03/18/2016 & 13:38h 07/27/2016 \\
				Capture end & 18:27h 06/26/2016 & 09:27h 08/29/2016\\
				Attacks start & N/A & 00:00h 07/28/2016 \\
				Attacks end & N/A  &  12:00h 08/09/2016\\
				Number of files & 17 & 6 \\
				Size (compressed) & 181GB & 55GB \\ 
				\# Connections & $\approx$ 13,000M & $\approx$ 3,900M\\[0.5ex]
				\hline
			\end{tabular}
	}}
	
\end{table}

The UGR'16 data set was used to evaluate MBDA in its original work~\cite{mbda}. This application of MBDA, following a completely unsupervised anomaly detection approach, showed high performance in the detection of attacks with exception to the NERISBOTNET. Later, the detection performance was improved by using a semi-supervised extension of MBDA~\cite{8628992} based on Partial Least Squares (PLS)~\cite{MC}\cite{GK86}. More recently, we showed that better performance than using semi-supervised methods can yet be achieved by properly performing outlier isolation in the background traffic~\cite{QIQO}. Importantly, comparing MBDA to the One-Class Support Vector Machine (OCSVM), a widely used black-box anomaly detection approach, we found that outlier isolation impacts by far more than the specific anomaly detection method. We will benchmark the performance of the feature learning approach proposed in this paper against all these previous results.

Intensive Big Data analysis requires a parallel computer. We used a multi-GPU DGX-1 server with dual 20-core processors (80 threads) and 512GB RAM. Python scripts using the FCParser run on top of the parallel hardware as grid jobs. The paralellization of the {upstream} phase, from learning to parsing, is straightforward. We can split data in parallel jobs in agreement with the data file partition (see Algorithm 1), and the result is simply appended. This approach can also be applied in the {downstream} phase. The analysis stage was performed with the MEDA Toolbox in a regular laptop.

{ Both the anonymized raw data and the corresponding feature data of UGR'16 can be found at \url{https://codas.ugr.es/animalicos/en/downloads.php}.}

\subsection{The Dartmouth Wi-Fi network Case Study}

Dartmouth College has a compact campus with over 200 buildings on 200 acres. The original evolution of the network is documented in the series of early papers~\cite{kotz:jcampus,henderson:jvoice,wifi}. The number of students, staff, and academic faculty reached near 6,500, 3,300 and 1,000, respectively, at the end of 2018, and the number of Access Points (APs) was above 3,000.  Researchers at Dartmouth have been capturing data about the usage of the network for many years, providing a perfect case study for tools like MBDA.

In this paper we analyze a data capture containing the connections of users to the network in the seven years: from 2012 to 2018~\cite{wifi}. { To collect the trace, the Dartmouth network operators configured the Cisco network controllers to forward a record of network activity to the research team's servers in the form of Simple Network Management Protocol (SNMP) traps~\cite{rfc1157} (see an example of trap in the Supplementary Materials). During the seven-year period, the network infrastructure (comprising Cisco network controllers and access points) was reasonably consistent.  The capture is thus a trace comprising a sequence of records (``traps''); each record includes a trap type (TT) and a set of fields labeled with object identifiers (OIDs). Anonymized data identifies who associated to the network with an anonymous tag, when the association took place, the (anonymized) device and APs involved in the connection and, thus, the approximate location and movement of each device and user throughout the capture. No traffic content is provided in the data.
}

The capture reveals the statistics in Table~\ref{tab:stats}. The data set contains a total of 5 Billion traps and 7 TB of data. A total of 38K authenticated users  and an undetermined number of non-authenticated users have been connected to the network in the last seven years, using 600K devices. 
The network infrastructure supports several SSIDs, primarily \emph{Dartmouth Secure}, the WPA2-Enterprise authenticated college network, \emph{Dartmouth Public}, a public-access network, and \emph{eduroam}, the world-wide roaming network for educational institutions~\cite{eduroam}.  Dartmouth Secure was entirely replaced by eduroam in the final months of the capture.

\begin{table}[htbp]
	\caption{Details of the SNMP trap capture at Dartmouth College.} \label{tab:stats}
	\centering{
		\small{
			\begin{tabular}{|l|c|}
				\hline \textbf{Statistic}
				& \textbf{Number}   \\
				\hline Capture period & Jan 1st 2012 - Dec 31st 2018 \\
				& (2556 days) \\
				log entries (SNMP traps) & 5 Billion \\ 
				Data Size (raw) & 7 TB \\
				Access points & 3,330 \\
				Authenticated Users & 38,096 \\
				Stations & 624,903 \\
				SSIDs & 20 \\ \hline
			\end{tabular}
	}}
\end{table}


We used the Anthill Compute Cluster hosted by the Computer Science Department at Dartmouth for both the {upstream} and the {downstream} phase. Again, the analysis stage was performed with the MEDA Toolbox in a regular laptop.

{ The anonymized raw data in JSON format can be provided by request to Prof. Kotz. The corresponding feature data can be found at \url{https://codas.ugr.es/animalicos/en/downloads.php}.


\subsection{Ethical statement}  

Raw data were anonymized following state-of-the-art practices, as explained in the original papers~\cite{macia2018ugr,wifi}. FAAC feature data, due to its nature, do not contain any personal information.

}

\section{UGR'16}
\label{sec:ugr}

Let us start with the application of the MBDA pipeline in the first case study. Our goal in this case is to automatically identify the attacks in the capture.

\subsection{{Upstream}}

\subsubsection{Feature learning}

We can think of at least two alternative ways to assess our approach for feature learning with the UGR'16 data set. One intuitive approach would be to learn the most prevalent features of background traffic in the training dataset, and then apply them for the detection of the attacks in the test set. This approach would render a purely unsupervised MBDA, equivalent to the work in~\cite{mbda}. Unfortunately, the background traffic also contains unlabeled anomalies and real attacks, and the evaluation based only in the artificial attacks may not be conclusive~\cite{QIQO}. An alternative and arguably more objective option is to learn the features from part of the flows corresponding to the artificial attacks themselves, and assess if they provide an improved performance in the detection of the remaining flows of those attacks. This is the choice we take in the paper, which corresponds to a semi-supervised MBDA approach similar to the one in~\cite{8628992}.  In a practical situation, the analyst would apply this approach when she wants to optimize the anomaly detector for specific (common) attacks. Still, given the unsupervised nature of the core of MBDA, MSNM, we retain the ability to detect unknown attacks.   

In agreement with the semi-supervised version of MBDA in~\cite{8628992}, we performed the feature learning on the raw files of the attacks corresponding to the first third of the test dataset, i.e., the first 4 days of attacks. The learning algorithm \texttt{fclearner.py} was launched in parallel in 24 processing jobs, one per hour in the day. 
The sampling interval, consistently with previous work, is set to 1 minute, and we set $T_l$ = 0.01 and $T_g$ = 0.001. Input variables were the source and destination port, the protocol and the tcp flags. Given the Netflow data is structured, the regular expressions of the variables are simply the location of the variable in the entries of the csv file with the raw dataset. 
The learning process resulted in a total of 396 features, most of them related to individual ports, and approximately 3 times the number of features in previous papers (134 manually selected features)~\cite{mbda,8628992,QIQO}. We also considered a second set of learned features obtained for $T_l = T_g$ = 0.01, which resulted in a subset of the first set with a total of only 17 features. The whole learning process using the parallel hardware took 33 hours. {
 In the training dataset we have a total of $4.8B$ of words. This represents a learning speed of $6.1\cdot 10^6$ words/CPUhour in parallel mode. As a reference, \emph{word2vec}~\cite{mikolov2013efficient} initial computations reflect a best case in parallelization of $6B$ words processed in 140 CPUs and 2 day time, that is, $8.9\cdot 10^5$ words/CPUhour. In the conclusions, authors report an optimized C++ multi-thread, single CPU, implementation that can provide speed in the order of $10^9$ words/CPUhour. Taking into account that C++ is between 1 and 2 orders of magnitude faster than Python~\cite{fourment2008comparison,zehra2020comparative}, and that $\texttt{fclearner.py}$ was not optimized for performance, it follows that a new package programmed on a faster language would be an interesting future contribution.}

\subsubsection{Parsing}

We use FCParser to generate feature vectors with two variants of configuration files learned from the data: with 396 and 17 features, respectively. In agreement with the learning phase, we consider feature vectors for intervals of one minute. This generates a total of approx. 160K observations of 396/17 features, which can be handled with the Big Data version of the MEDA Toolbox in a regular computer~\cite{mbda}. Recall that we can vary the level of detail by using different time resolutions: if we use 1 hour intervals rather than 1 minute intervals, the number of observations would be reduced 60-fold to approx. 2,7K, but the resolution of detection would also be reduced. 

The parsing was parallelized again in 24 processing jobs, one per hour in the day, and the whole process took 20 days. While this is a lengthy process, considering that the trace corresponds to 4 months of data, we can conclude that the parsing approach can be implemented in real time. In any case, this time can be reduced using a larger computer and properly arranging the input data for parallelization (see Algorithm 1). 

\subsection{Analysis}

We focus on the ability of MBDA to identify the attacks in the part of the test set not used for the feature learning, that is, the last 8 days. To benchmark the anomaly detection performance with previous results, we compute the false positive rate (FPR) and true positive rate (TPR) of detection, and in turn the Receiver Operating Characteristic (ROC) curves, that shows the evolution of the TPR versus the FPR for different values of the anomaly detection threshold. A practical way to compare several ROC curves is with the Area Under the Curve (AUC), a scalar that quantifies the quality of the anomaly detector. The anomaly detector should present an AUC as close to 1 as possible, while an AUC {around} 0.5 corresponds to a random classifier.

\begin{figure*}[!ht]
	\centering
	\subfigure[]{\includegraphics[width=0.4\textwidth]{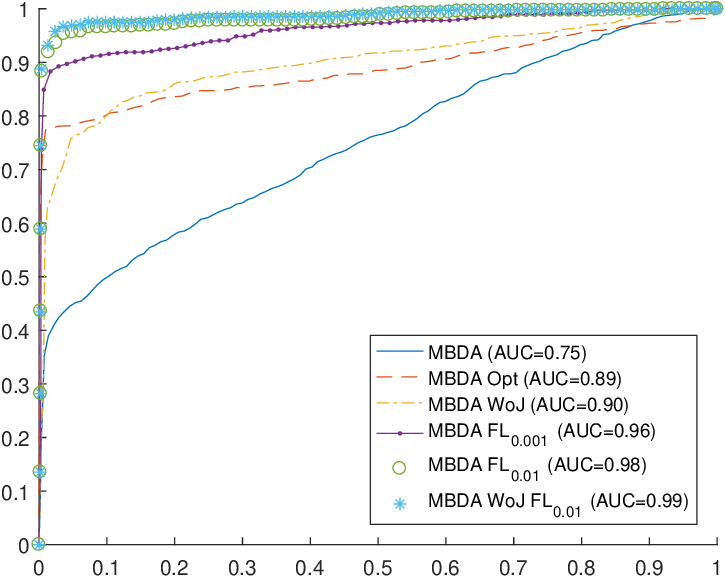}} \subfigure[]{\includegraphics[width=0.4\textwidth]{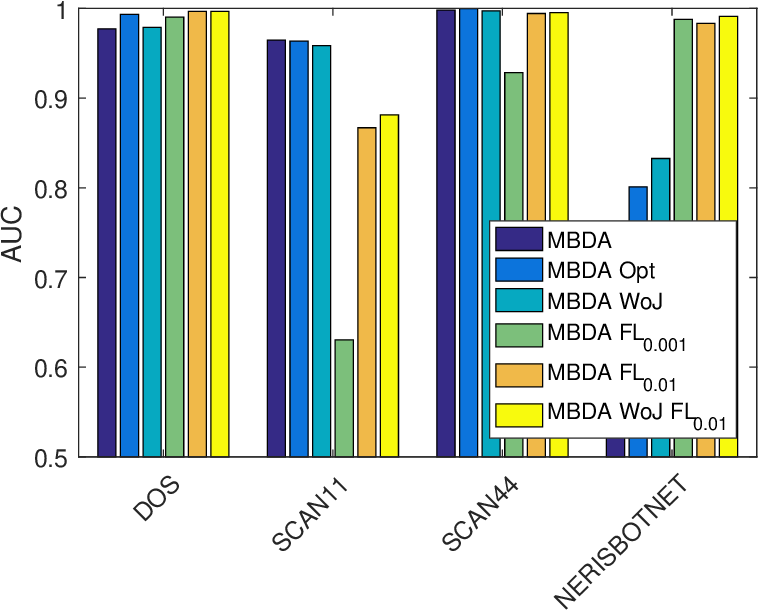}} 
	\caption{ROC curves (a) and attack-type based AUC results (b) for a set of different solutions based on MBDA.} \label{ROC}
\end{figure*}

Figure \ref{ROC} shows the comparison of a number of different MBDA variants (see also Table \ref{tab:MBDAvars}), including:
\begin{itemize}
	\item MBDA: The original, unsupervised approach~\cite{mbda} trained with the complete training dataset using manually selected features.
	\item MBDA Opt: The semi-supervised extension of MBDA~\cite{8628992}
 	trained with the complete training dataset using manually selected features and optimized with Partial Least Squares (PLS) using the attacks of the first four days of the testset.
	\item MBDA WoJ: The unsupervised MBDA with manually selected features trained without June, where an anomaly with a similar pattern as a botnet was found~\cite{QIQO}.
	\item MBDA FL$_{0.001}$: The semi-supervised MBDA trained with the complete training dataset and with the 396 features learned for $T_g$ = 0.001 using the attacks of the first four days of the testset.
	\item MBDA FL$_{0.01}$: The semi-supervised MBDA trained with the complete training dataset and with the 17 features learned for $T_g$ = 0.01 using the attacks of the first four days of the testset.
	\item MBDA WoJ FL$_{0.01}$: The semi-supervised MBDA trained without June and with the 17 features learned for $T_g$ = 0.01 using the attacks of the first four days of the testset.
\end{itemize}

\begin{table*}[t]
	\caption{MBDA variants under study.}
	\label{tab:MBDAvars}
	\centerline{
		\small{
			\begin{tabular}{l c c c}
				\hline \\[-1.5ex]
				{\bf Name} & {\bf Type} & {\bf Features} & {\bf June in training data} \\[0.5ex]
				\hline \\[-1.5ex]
				MBDA & unsupervised & manual & Yes \\
				MBDA Opt & semi-supervised & manual & Yes \\
				MBDA WoJ & unsupervised & manual & No \\
				MBDA FL$_{0.001}$ & semi-supervised  &  learned ($T_g$ = 0.001) & Yes \\
				MBDA FL$_{0.01}$ & semi-supervised  &  learned  ($T_g$ = 0.01) & Yes \\
				MBDA  WoJ FL$_{0.01}$ & semi-supervised  &  learned  ($T_g$ = 0.01) & No \\[0.5ex]
				\hline
			\end{tabular}
	}}
	
\end{table*}

Figure \ref{ROC}(a) presents the general ROC curves, obtained for the four types of attacks, and Figure \ref{ROC}(b) represents the AUCs per attack type, { where each AUC is computed
comparing normal data with the specific type of attack, leaving out the observations corresponding to the other attack patterns.}

Our proposal for feature learning generally outperforms other methods based on manually selected features. We can see that the improvements are mainly on NERISBOTNET attacks, while the performance for SCAN attacks is generally better for versions of MBDA with manually selected features.

We can obtain more information about the root causes for aforementioned performance differences among the models when detecting specific attacks. For that purpose, we use the approach presented in~\cite{QIQO} that combines diagnosis plots~\cite{FUENTESGARCIA2018194}), univariate box plots and t-tests for statistical inference. We compare MBDA Opt and MBDA FL$_{0.01}$ as representatives of models with manual features and learned features, respectively, since both are semi-supervised and trained with the complete training dataset. 

\begin{figure}[!ht]
	\centering
	\subfigure[]{\includegraphics[width=0.6\columnwidth]{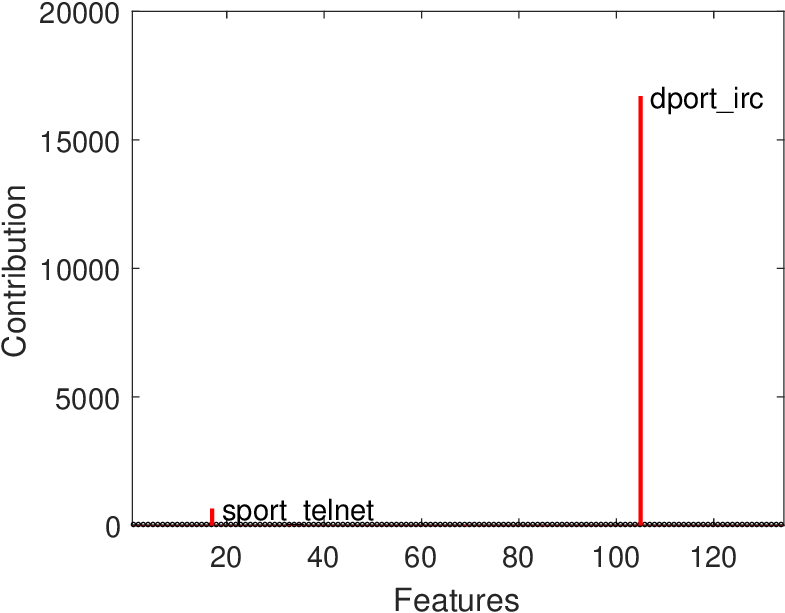}} \subfigure[]{\includegraphics[width=0.6\columnwidth]{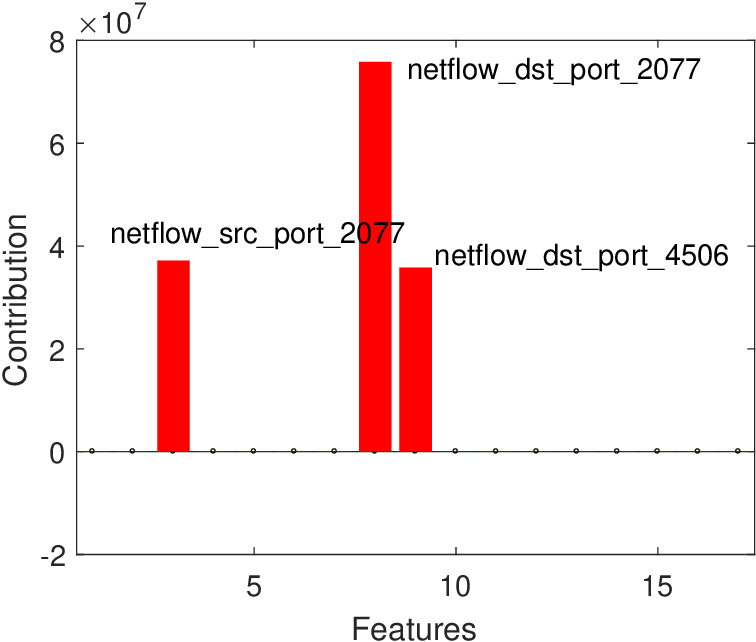}} 
	\caption{Profile of detection of NERISBOTNET attacks with MBDA Opt (a) and MBDA FL$_{0.01}$ (b) using oMEDA.} \label{NERIS}
\end{figure}

The diagnosis plots for NERISBOTNET attacks are shown in Figure \ref{NERIS}. { Diagnosis plots are obtained by comparison of the anomaly with the normal data using a specific model (MBDA Opt and MBDA FL$_{0.01}$). It is represented as a bar diagram of the features. Positive bars identify features in which the anomaly has higher value than normal data, and negative bars represent the opposite (only positive bars are found in the figure).} MBDA Opt emphasizes irc and telnet ports, while MBDA FL$_{0.01}$ focuses on ports 2077 and 4506\footnote{The \texttt{fclearner.py} tool combines the label of the variable with the regular expression learned to create the label of a feature. This is why we only see numbers in the labels, unlike in manual features.}. All selected features yield statistically significant differences between background traffic and NERISBOTNET attacks, as illustrated in Figure \ref{NERIS2}{, which shows boxplots and t-tests significance results between normal data (Neg) and the attacks (Pos)}. However, according to AUC results, learned features provide a more powerful detection.

\begin{figure}[!ht]
	\centering
		\subfigure[Ttest $\mathbf{p_-value < 0.01}$]{\includegraphics[width=0.4\columnwidth]{./Figures/BoxNeris\_dport\_irc}} \subfigure[Ttest $\mathbf{p_-value < 0.01}$]{\includegraphics[width=0.4\columnwidth]{./Figures/BoxNeris\_netflow\_dst\_port\_2077}}  
		\subfigure[Ttest $\mathbf{p_-value < 0.01}$]{\includegraphics[width=0.4\columnwidth]{./Figures/BoxNeris\_netflow\_src\_port\_2077}}  
		\subfigure[Ttest $\mathbf{p_-value < 0.01}$]{\includegraphics[width=0.4\columnwidth]{./Figures/BoxNeris\_netflow\_dst\_port\_4506}} 
	\caption{Boxplots and ttests of selected features in background traffic (Negative) versus NERISBOTNET traffic (Positive).} \label{NERIS2}
\end{figure}

\begin{figure}[!ht]
	\centering
	\subfigure[]{\includegraphics[width=0.6\columnwidth]{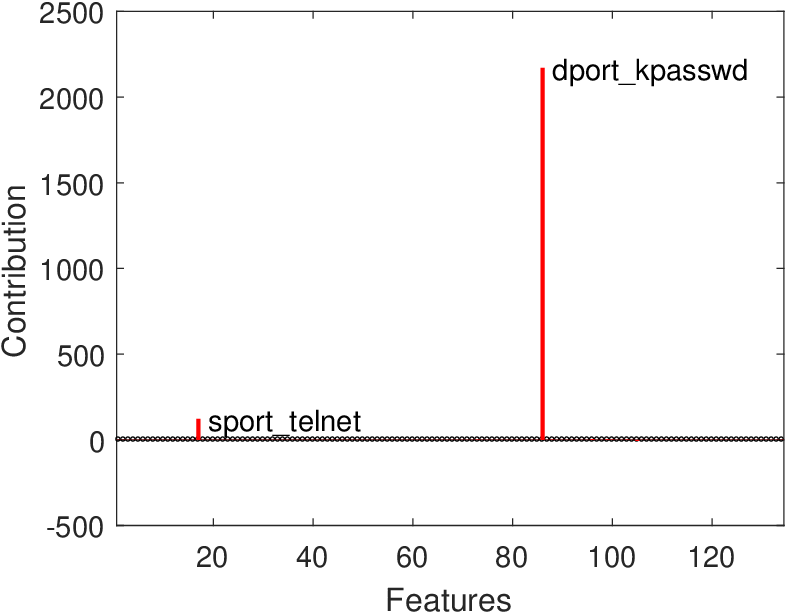}} \subfigure[]{\includegraphics[width=0.6\columnwidth]{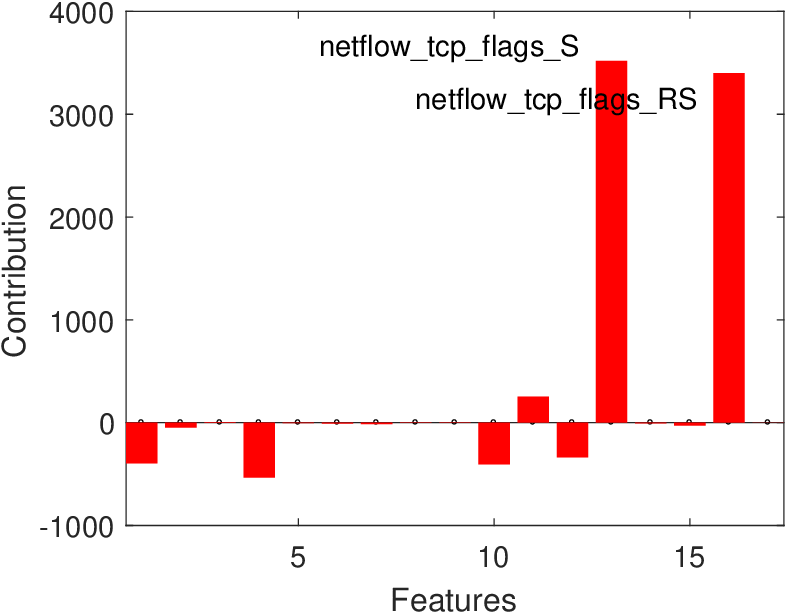}} 
	\caption{Profile of detection of SCAN11 attacks with MBDA Opt (a) and MBDA FL$_{0.01}$ (b) using oMEDA.} \label{SCAN}
\end{figure}

We repeat the same procedure for SCAN11 attacks, shown in Figures \ref{SCAN} and \ref{SCAN2}. MBDA Opt  emphasizes kpasswd and telnet ports, while MBDA FL$_{0.01}$ focuses on the TCP flags, in particular Sync and the combination of Reset and Sync. In this case, the manual selection of features provides a more powerful detection in terms of AUC. However, neither pattern of detection is perfect: in SCAN attacks, the attacker sends probing messages to find open ports, and does that for a large number of different ports. MBDA Opt only detects the attack because there is one single port of those tested, kpasswd, with negligible background traffic, but the diagnosis does not reflect the true pattern of attack. MBDA FL$_{0.01}$ provides limited performance because the learning approach based on prevalence and counting features cannot capture the pattern of the attack. Future work may look at different learning loss functions other than prevalence and/or alternative definitions of features that capture distributional information of a variable, like the number of different ports in a time interval.     

\begin{figure}[!ht]
	\centering
	\subfigure[Ttest $\mathbf{p_-value < 0.01}$]{\includegraphics[width=0.4\columnwidth]{./Figures/BoxSCAN11\_dport\_kpasswd}} \subfigure[Ttest $\mathbf{p_-value < 0.01}$]{\includegraphics[width=0.4\columnwidth]{./Figures/BoxSCAN11\_sport\_telnet}} \subfigure[Ttest $\mathbf{p_-value < 0.01}$]{\includegraphics[width=0.4\columnwidth]{./Figures/BoxSCAN11\_netflow\_tcp\_flags\_S}} 
	\subfigure[Ttest $p_-value = 0.83$]{\includegraphics[width=0.4\columnwidth]{./Figures/BoxSCAN11\_netflow\_tcp\_flags\_RS}} 
	\caption{Boxplots and t-tests of selected features in background traffic (Negative) versus SCAN11 traffic (Positive).} \label{SCAN2}
\end{figure}

\begin{figure}[!ht]
	\centering
	\includegraphics[width=0.4\textwidth]{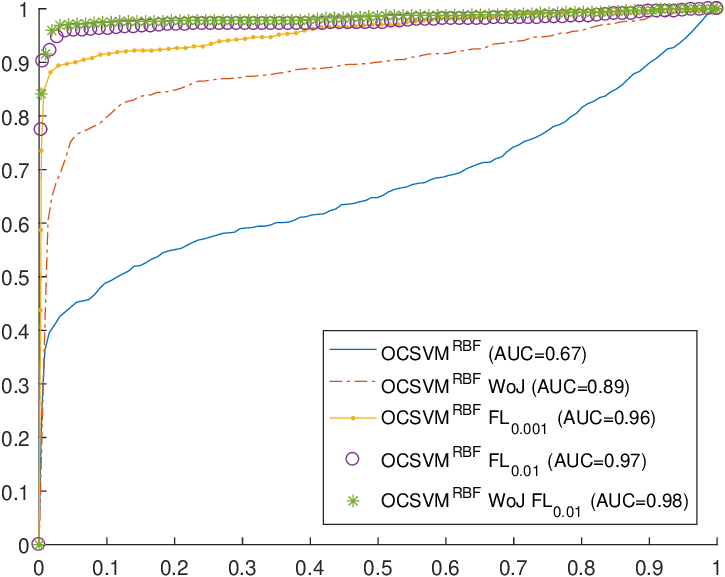}
	\caption{ROC curves for OCSVM variants.} \label{ROCSVM}
\end{figure}

{A reasonable question would be if the learned features with our proposal are only useful for MBDA or they can be used in combination with other Machine Learning approaches. Following previous work~\cite{QIQO}, we assess the performance of the one-class support vector machine (OCSVM)~\cite{Scholkopf2000,Scholkopf2001} based on radial basis functions (RBF), the most extended kernel choice, with manual and automatically learned features. OCSVM is a non-linear tool, and therefore has the advantage over MBDA to model non-linear behavior in the model of normal traffic, but it does not (in principle) have the same capability to handle highly multivariate feature data as MBDA. Thus, both methods have very different nature. The results are presented in Figure \ref{ROCSVM}. We can see that the performance of OCSVM is very similar to that of MBDA, and it significantly improves with the automatically learned features. This is an interesting result, since even a blackbox model like OCSVM with RBF can get useful explanations with new interpretation tools like SHAP~\cite{NIPS2017_8a20a862}.}

\subsection{{Downstream}}

The previous analysis compared the accuracy of detection at time interval (1 minute) level. As an illustrative example of the {downstream} step, we compare here the accuracy of detection of the NERISBOTNET attack at flow-level by MBDA Opt and MBDA FL$_{0.01}$. Results are presented in Table \ref{tab:upstream_UGR16} in terms of the number of true positives (TP) and negatives (TN), the number of false positives (FP) and negatives (FN), the accuracy ((TP+TN)/Total) and the False Discovery Rate (FDR = FP /(TP+FP)). While accuracy levels are close to 1.00, like those reported earlier~\cite{mbda}, the FDR is a more relevant statistic to assess the difficulty in the process of root cause analysis. The FDR gives us an estimate of the relative number of false alarms an analyst will have to face in the process of alarm validation. In the example, we can see that the MBDA based on feature learning reduces the relative number of false alarms to only 1.8\%, which is a competitive statistic and much lower that the one using manual features.    

\begin{table*}[!ht]
	\caption{Comparison of MBDA Opt and MBDA FL$_{0.01}$ in the {downstream} step in the detection of the NERISBOTNET attack. We present the number of true positives (TP) and negatives (TN), the number of false positives (FP) and negatives (FN), the accuracy ((TP+TN)/Total) and the False Discovery Rate (FDR = FP /(TP+FP)). The total number of flows in the test data (the part used for performance evaluation, that is, the last 8 days) is 1,074,221,524, and the number of attack flows in the same data is 1,074,493 (0.1\% of the total).}
	\label{tab:upstream_UGR16}
	\centering{
		\small{
			\begin{tabular}{|l|l|l|l|l|l|l|}
				\hline
				\textbf{Method} & \textbf{TP} & \textbf{TN} & \textbf{FP} & \textbf{FN} & \textbf{Accuracy} & \textbf{FDR} \\
				\hline
				MBDA Opt          & 33,613 & 1,073,101,984 & 45,047 & 1,040,880 & $\approx 1.00$ & 0.570 \\
				MBDA FL$_{0.01}$ & 61,261 & 1,073,145,928 & 1,103 & 1,013,232 & $\approx 1.00$ & 0.018 \\
				\hline
			\end{tabular}
	}}
\end{table*}

{It should be noted that while the FDR is adequately low for MBDA FL$_{0.01}$, the number of FN is still very high (yet lower than in the case of the manually based MBDA), which renders the flow-level sensitivity of the method undesirably low. A low FDR with a low sensitivity means that the system will provide the security analyst with a list of alarms that only contains a subset of security-relevant flows, but that most flows in the list are going to be truly relevant in terms of security. While this means that there is still margin for improvement, this is indeed a promising result considering that most industrial security appliances (e.g., SIEMs or IDSs) are severely affected by false alarms, which reduce their practical usefulness. On the other hand, we previously found in Figure \ref{ROC} that the MBDA approach based on feature learning is very accurate at time interval-level (both in terms of sensitivity and specificity), which leads to conclude that most real threats are detected by the system, but only a small subset of related flows are recovered in the Downstream.}

\section{Dartmouth Wi-Fi}
\label{sec:action} 

Let us move on to the analysis of the Dartmouth Wi-Fi capture. Our goal here is to visualize and understand the main factors of variance in the connection data.

\subsection{{Upstream}}

\subsubsection{Feature learning}
\label{sec:learning2}

\begin{table}[!ht]
	\caption{First 10 SNMP OIDs with more prevalence in the data. CLAM refers to CISCO-LWAPP-AP-MIB, AWM to AIRESPACE-WIRELESS-MIB and CLDCM to CISCO-LWAPP-DOT11-CLIENT-MIB. }
	\label{tab:vars2}
	\centering{
		\small{
			\begin{tabular}{|l|l|l|}
				\hline
				\textbf{Label} & \textbf{Type} & \textbf{Presence} \\
				\hline
				CLAM::cLApDot11IfSlotId & OID &	0.45\\
				AWM::bsnAPName & 	OID &			0.41\\
				AWM::bsnStationMacAddress & 	OID &			0.39\\
				AWM::bsnStationAPIfSlotId & OID &				0.39\\
				AWM::bsnStationAPMacAddr & OID &				0.39\\
				AWM::bsnStationUserName & OID &				0.39\\
				CLAM::cLApName & OID &				        0.36\\
				CLDCM::cldcClientMacAddress &	OID &			0.22\\
				CLDCM::cldcApMacAddress &	OID &			0.22\\
				AWM::bsnDot11StationAssociate &	TT &			0.20\\
				\hline
			\end{tabular}
	}}
\end{table}

We performed the learning strategy in two steps to identify high variance features in the Wi-Fi data. First, the learning algorithm \texttt{fclearning.py} was launched in parallel in 2556 processing jobs, each one for a different day in the capture, using a sampling interval of 1 day and a threshold values $T_l = 0.05$ and $T_g = 0.01$. Input variables were the regular expressions for a SNMP object identifier (OID) and for the trap type (TT) (see Supplementary Materials for more detail). The output is 2556 configuration files, one per day, with the set of most prevalent OIDs and TTs in each day. That way, we learn as features all those OIDs or TTs with a daily prevalence above the 5\% in at least one day and a total prevalence above 1\%. This resulted in a total of 90 features, including prevalent OIDs, TTs and default features. The ten most prevalent features are shown in Table~\ref{tab:vars2}, where we make the distinction between OIDs representing trap types (TTs) and the rest. 

The whole learning process using the parallel hardware and multi-threading (4 threads per processor) took 12 hours, during which a maximum of 150 jobs were processed in parallel. This means that the processing time could be further reduced 17-fold using a larger computer cluster, where as many as 2556 jobs could be run in parallel. {
 Computing the number of words is more tricky in this case. Considering that each field in a trap has an average of 50 characters, we can estimate a total of 140B words. Considering that each trap has an average of 20 fields, we obtain an estimate of 100B words, which is reasonably close. Using the lowest estimate, the learning speed is of $5.6\cdot10^7$ words/CPUhour, again competitive in computational time for a non-optimized Python tool.}

\begin{figure*}[!ht]
	\centering
	\subfigure[]{\includegraphics[width=0.4\textwidth]{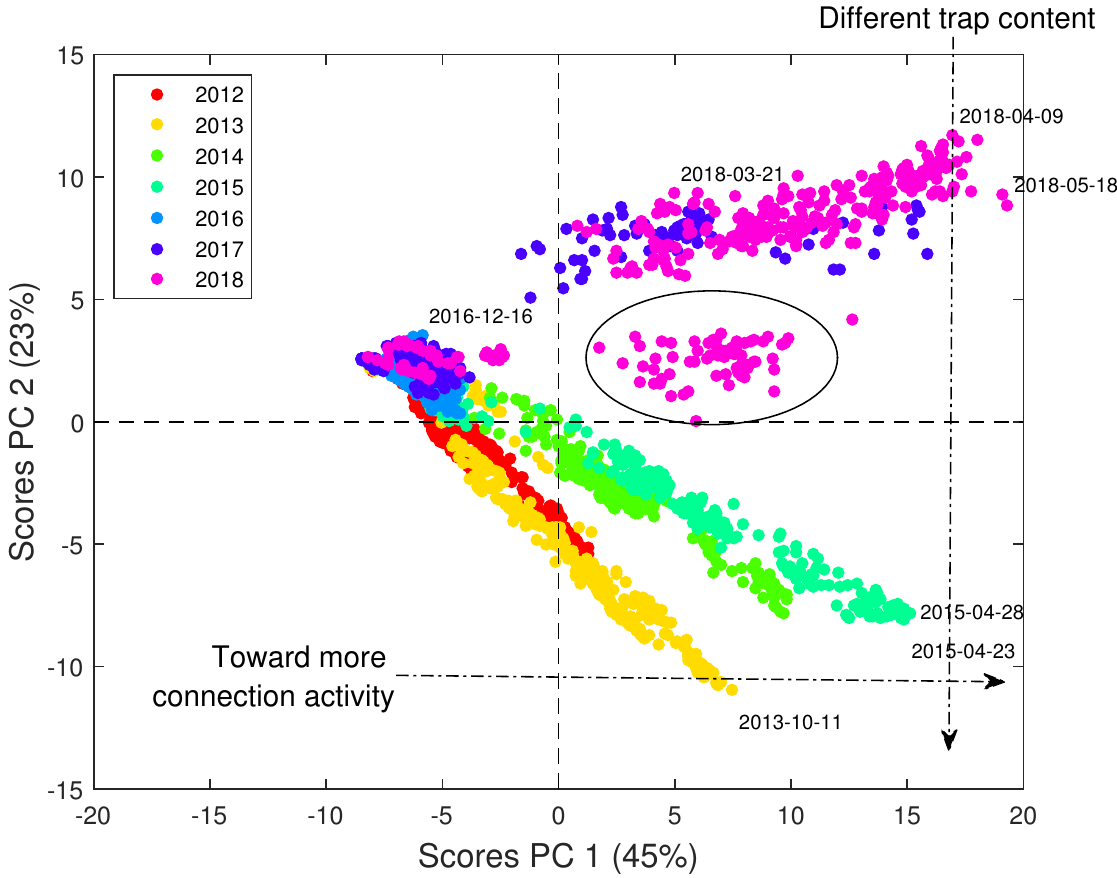}} \subfigure[]{\includegraphics[width=0.4\textwidth]{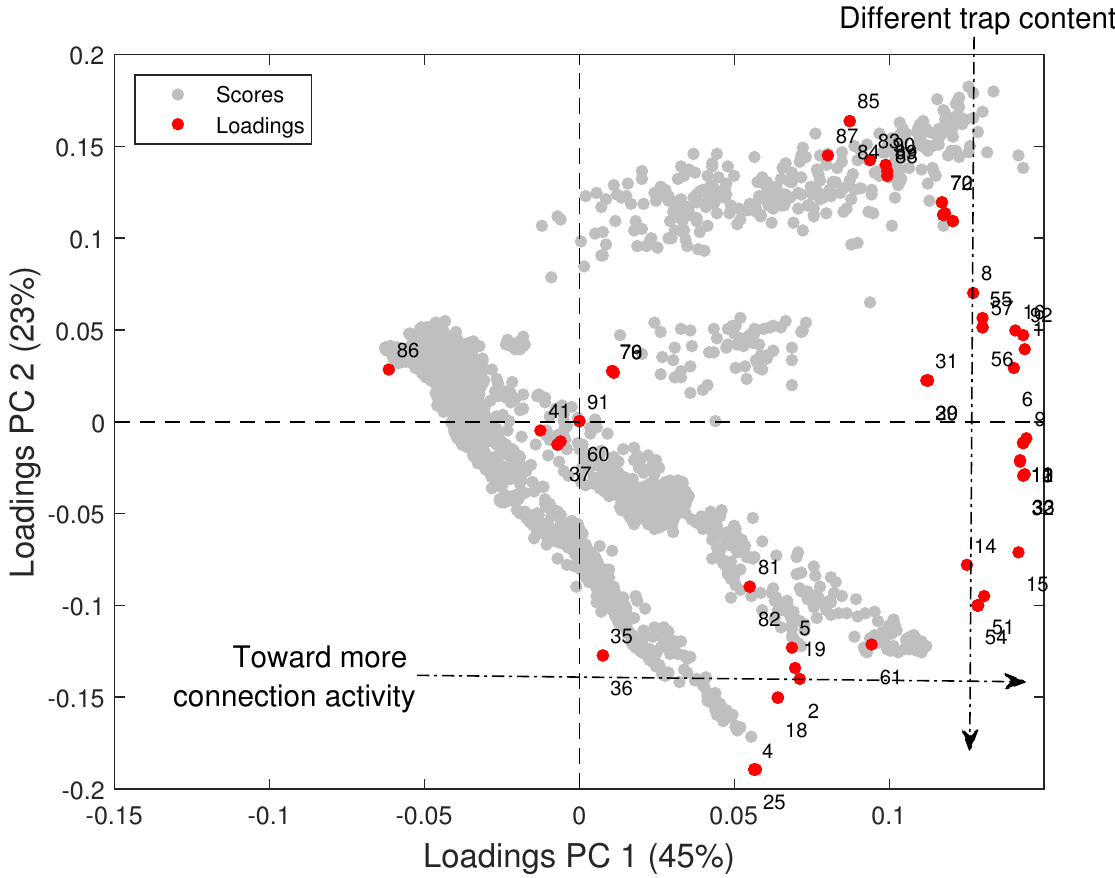}} 
	\caption{PCA scores and loadings for PC1 vs PC2.}
	\label{fig:scores}
\end{figure*}

\subsubsection{Parsing}

We use the FCParser to generate the feature vectors with the {aforementioned} configuration file learned from the data. In agreement with the learning phase, we consider feature vectors for intervals of one day. To the set of 90 learning features, we added the total number of traps and OIDs per day. 
This results in a compression of the data from 7TB to less than 1MB, yielding 2556 observations (days) of 92 features each {in matrix $\mathbf{X}$}. The compression conveniently transforms a Big Data set into a handleable data set in a common computer. Again, we can vary the level of detail by using different time resolutions or number of features.

The parsing was parallelized in 2556 processing jobs, one per day, and the whole process using the Anthill Computer Cluster and multi-threading (with a maximum of 150 jobs) took 15 hours. The resulting feature data is available on request from the authors.

\subsection{Analysis}
\label{sec:Detaction}



\subsubsection{Analysis with PCA}

Figure~\ref{fig:scores} depicts the plots corresponding to the first two PCs {in matrix $\mathbf{X}$} (refer to Supplementary Materials for other patterns found in subsequent PCs). Recall that matrix $\mathbf{X}$ contains 2556 rows, representing days of the capture, and 92 features. We present the score plot at the left of the figure and a bi-plot at the right. 
In the score plot, points represent the 2556 days of the data capture and are colored according to the year. In the bi-plot, (red) points represent the 92 features and the (gray) shadow represent the scores.

The first two PCs represent 68\% (45\% + 23\%) of the variance in the data. Because variance is a measure of the degree of change 
within the data set, these two PCs show the main patterns of change in $\mathbf{X}$. As a matter of fact, a variance of $68\%$ roughly indicates that only 1/3 of the patterns of change in the data is missing in this plot, giving an idea of how powerful PCA is for visualization. 

The score plot at Figure~\ref{fig:scores}(a) shows that the dots (days) with different colors are in different locations. This means that they are different in content, from which follows that there are large differences in prevalence of OIDs in different years. 

To interpret the bi-plot at Figure~\ref{fig:scores}(b), recall that the location of an observation (a day) will approach more the location of a feature (which represents counts of a specific OID) as the value of that feature increases in the observation. Thus, days with a large content on specific OIDs will be located closer in the plot to the loading representing that OID. The bi-plot shows that a large majority of the features  are located far from the center of coordinates towards the right side. Therefore, any day toward the right in the score plot will have a generally higher content of OIDs. Thus, as we traverse from left to right in the score plot, the days will have more connection activity. Busy periods are represented towards the far right of the plot, and vacations are clustered to the left, and we could say that the first PC (the horizontal direction in the score and loading plots) represents the general activity in the network. We annotated this in both plots using a horizontal arrow.

The bi-plot in Figure~\ref{fig:scores}(b) also shows that the variables are distributed from the bottom to top, and we see a similar distribution for the different years in the score plot: the first two years  are in the bottom and the last two in the top, with middle years in between. We also see a separated cluster of days in 2018, highlighted with a circle. A closer look reveals that all the days in the cluster belong to the period from September to November, when eduroam replaced Dartmouth Secure. The vertical pattern in the loading and score plots shows that the distribution of traps has changed across the years: days towards the top have a higher content of traps and OIDs represented by the features in the top and less of those in the bottom, and vice-versa. Again, we annotated this in the score plot and the bi-plot using a vertical arrow. Questioned about this difference, the network operators replied that there was an update in the controllers' software, which changed the types of SNMP traps that were collected. This variability in traps for different temporal periods makes the analysis of the data a real challenge. {If OIDs prevalence changes over time and we manually select a subset of OIDs as features by screening limited portions of the massive data, we may arrive at a different selection of features depending on which period of the data we visualize. Furthermore, the traps variability may go unnoticed if we manually select features that are only prevalent in one specific period. Our automatic learning approach solves this issue.} 


Regarding processing burden, the analysis performed in this section is completely interactive in a regular computer, meaning that the time to obtain each of the plots is on the order of seconds. 

\subsubsection{Analysis with MSNM}

\begin{table*}[htb]
	\centering
	\footnotesize
	\caption{Pre-diagnosis of the excursions of 2013 and 2017 with oMEDA.}
	\label{table:diag1}
	\begin{tabular}{c c }
		
		\hline
		\hline
		Timestamps& Features selected 
		\\
		\hline		
		2013-12-14 -- 2013-12-16& bsnDot11StationAuthenticateFail, bsnAuthenticationFailure, bsnDot11StationAssociateFail,\\
		&  bsnStationReasonCode, bsnAuthFailureUserType, bsnAuthFailureUserName   \\
		\hline		
		2017-10-16 -- 2017-10-30& ciscoLwappApIfUpNotify, ciscoLwappApIfDownNotify
		\\
		& cLApAdminStatus, cLApSysMacAddress, cLApPortNumber   \\
		\hline
	\end{tabular}
\end{table*}

\begin{figure}[!t]
	\centering
	\includegraphics[width=0.5\textwidth]{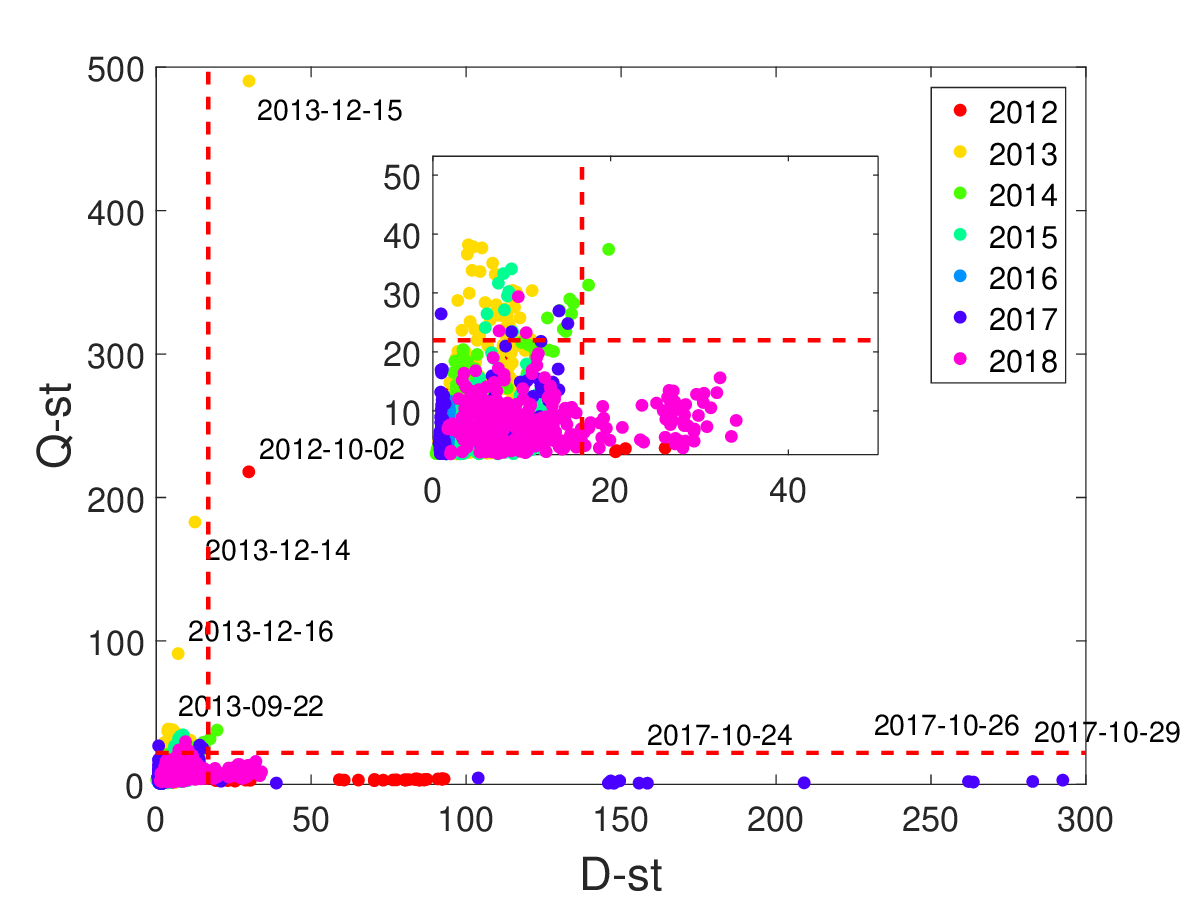}
	\caption{Multivariate Statistical Network Monitoring (MSNM) plot: D-statistic vs Q-statistic. At the top, zoomed image of the bottom left corner. }
	\label{fig:mspc}
\end{figure}

After the inspection of score and loading plots, one can visualize a summary of the whole data distribution in one single plot using MSNM: a scatter plot of the observations in terms of the D-statistic and the Q-statistic. 
The MSNM plot for the Wi-Fi data is shown in Figure~\ref{fig:mspc}. Anomalies are expected to surpass any of the two control limits: the vertical one for the D-statistic or the horizontal one for the Q-statistic. This plot is optimized for anomaly detection. Note that with only one visualization, the operator can identify the main patterns of change in 7TB of data. Clearly, in the plot we miss other details, like yearly and seasonal patterns and the difference in trap contents.
A main advantage of this plot is that it also includes residuals, containing  the remaining 6\% of the variance that is not accounted for in the six PCs (including those shown in the Supplementary Materials). The Q-statistic, which comprises a summary of the residuals, clearly identifies anomalies in 2012 and 2013, while the D-statistic finds several anomalous intervals in 2012 and 2017.

Note that the parsing (and thus of the learning process) has a principal impact in the visualization and anomaly detection of MBDA. For instance, we can  detect anomalies (e.g., excursions) only at the day level when using one-day resolution. If we want to detect anomalies in other time resolutions, we can modify the parsing configuration and re-run the {upstream} stage. Furthermore, differences in OID content can only be directly visualized if we include features for those OIDs (recall default features will still represent such differences to a certain level). Therefore, learning features of high variance is paramount to obtaining accurate insights of the data distribution.

\begin{figure}[t]
	\centering
	\subfigure[Excursion in 2013]{\includegraphics[width=0.4\textwidth]{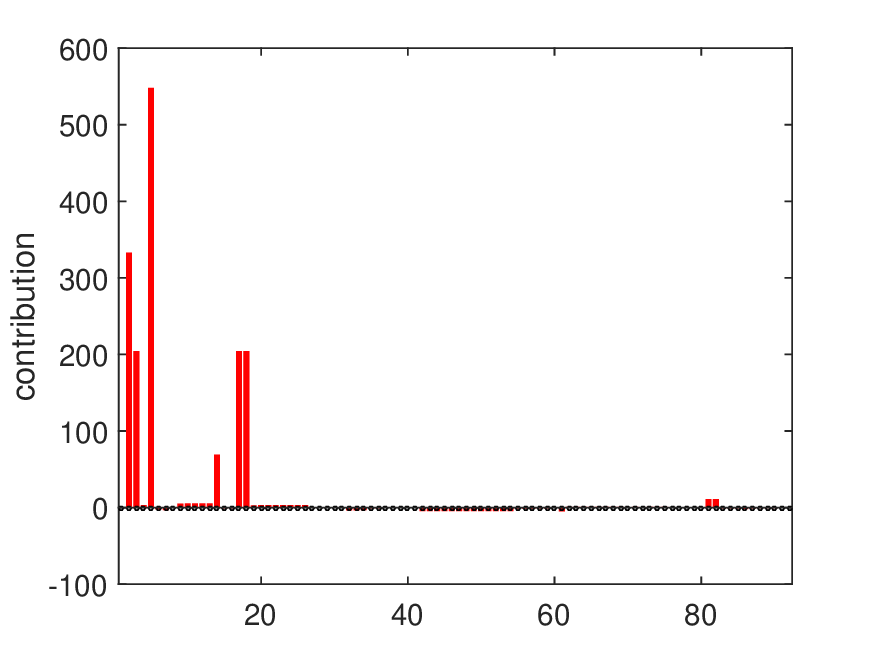}}
	\subfigure[Excursion in 2017]{\includegraphics[width=0.4\textwidth]{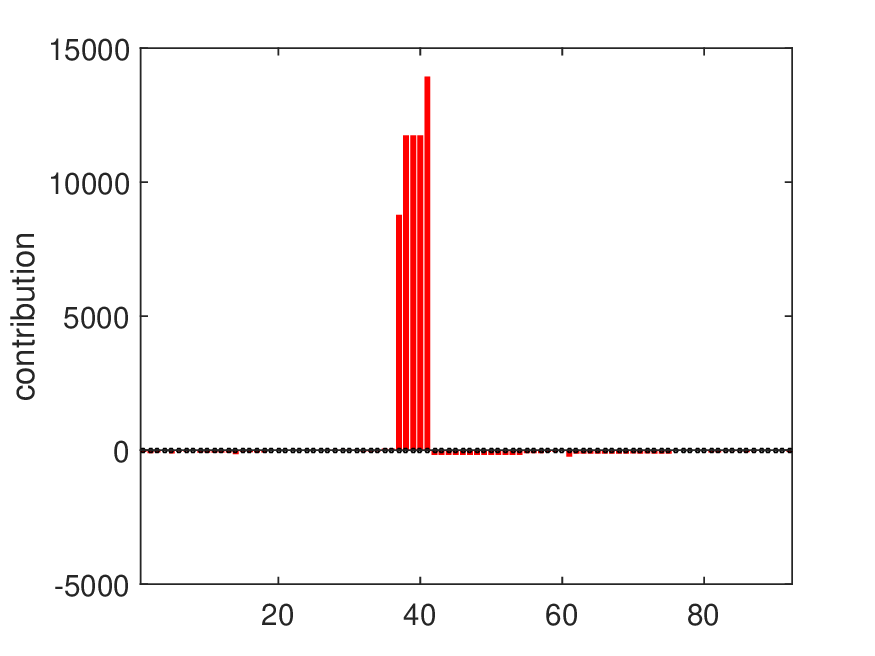}}
	\caption{Pre-diagnosis of the anomalies in 2013 and 2017 with oMEDA.}
	\label{fig:omeda}
\end{figure}

To illustrate the use of oMEDA in the diagnosis, we selected the anomalies in 2013 and 2017, which we found to be the main outliers in the Q-statistic and in the D-statistic, respectively. The plots are shown in Figure~\ref{fig:omeda}. The high bars identify the features that make the anomalous intervals different to the normal days. Each of the intervals are related to a different set of features. Table~\ref{table:diag1} lists the specific features. {We can see that the anomalies found are multivariate, since they are connected to multiple features, and it is this multivariate pattern that allows a better diagnosis of the anomaly.} We determined that the first anomaly (2013) is related to a large number of Authentication Fails, which in a subsequent analysis (not shown) we determined these fails were one order of magnitude higher than usual during the detected anomalous interval. The second anomaly (2017) is related to an unprecedentedly high number of re-starts of APs, two orders of magnitude higher than usual.

As for 2018, the network operators did not have any records of these old anomalies, but they suggested that the second one could be related to the installation of a security patch after the publication of a vulnerability. Effectively, October 16th of 2017, the famous KRACK attack against WPA2~\cite{Vanhoef:2017:KRA:3133956.3134027} and the corresponding patch was released to the public. Even if a restart is necessary after a patch installation, the number and duration (15 days) of the event is remarkable, evidencing that a major management problem took place that went unnoticed into the massive stream of SNMP traps.

Like the PCA analysis, the MSNM analysis is fully interactive and easily done in a regular computer.

\subsection{{downstream}}
\label{sec:Deparsing}

\begin{table*}[htb]
	\centering
	\footnotesize
	\caption{Deparsing of the anomalies of 2013 and 2017 with oMEDA.}
	\label{table:diag}
	\begin{tabular}{c c  c c c }
		
		\hline
		\hline
		Timestamps&  log entries/tot  & \#APs & \#Stations & \#Users  \\
		\hline		
		2013-12-14 -- 2013-12-16 &      5.4M/8.4M (64\%)  &  824 & 595 & 103 \\
		\hline		
		2017-10-16 -- 2017-10-30 & 19.0M/64.1M (30\%)         &    1,376    &  0  & 0  \\
		\hline
	\end{tabular}
\end{table*}

We applied the {downstream} stage with the FCParser to the anomalies in 2013 and 2017 in the Wi-Fi data set. We parallelized the processing using the Anthill Computer Cluster and multi-threading (4 threads per processor), with as many parallel jobs as days in the excursions. The first anomaly took 30 minutes to be processed, and the second one 135 minutes. The output is a file per anomaly, containing the traps involved, which represent a subset of total set of traps in the corresponding periods of time. Table~\ref{table:diag} provides some statistics of the deparsing. The human operator can use the output files to retrieve more information about the anomalies, like the main actors (APs, users, devices) involved.

{
\section{Conclusion}
\label{sec:conclusion}

In this paper, we introduce a new feature learning approach especially suited for the Multivariate Big Data Analysis (MBDA), an interpretable data analysis tool optimized to analyze Big Data streams. MBDA has shown high capability for anomaly detection, diagnosis and network data understanding, but its application to Big Data problems is limited by the fact that it requires a manual definition of the data features. In this work, we overcome this limitation by proposing an approach for automatic learning of interpretable features that is shown to improve the performance of MBDA while maintaining its interpretability. Our learning approach is demonstrated in two real case studies: a Netflow trace from a TIER-3 ISP and a connection trace from a campus Wi-Fi network. The results illustrate that the tandem of feature learning and MBDA can bring light into massive data sets for network-monitoring purposes. 
}

{ Unlike alternative feature learning approaches, like Word2Vec, our method is centered on interpretability. Furthermore, we learn features using prevalence as our main optimization criteria.	As a result, our approach is not particularly optimized for regression, classification or anomaly detection tasks. However, we show in the ISP case study that the method can be leveraged for such a purpose (in our case anomaly detection, but extensions to regression or classification are straightforward) by properly choosing the training data from which we obtain the features, and that performance improvements can be extended to Machine Learning methods other than MBDA. However, a major contribution of the combination of our proposal with MBDA is the possibility to visualize massive datasets in a sensible and interpretable way, as we show in the campus Wi-Fi case study. 	
	
Our proposal, although promising, is far from perfect. We found that the integration of feature learning with MBDA can provide very accurate detection of anomalous events, and the diagnosis can recover related records with high specificity, but the sensitivity at record level can still be much improved. Furthermore, the diagnosis is only accurate when suitable features for that diagnosis are learned, for which prevalence may not be the optimal criteria and other learning criteria may be studied. Finally, it should be noted that the proposed learning algorithm can only provide meaningful individual features, but it is not capable to yield grouping features, e.g., a feature that counts the prevalence of a group of destination ports or a subnet of IPs. Future work on these topics can be an interesting means to improve numerical and interpretational performance.}

%
\section*{Acknowledgement}
\label{sec:Acknowledgments}

This work was supported by Dartmouth College, and in particular by the many network and IT staff who assisted us in configuring the Wi-Fi network infrastructure to collect data, and who patiently answered our many questions about the network and its operation.
We furthermore appreciate the support of research colleagues and staff who have contributed to our data-collection and data-analytics infrastructure over the years: most notably Wayne Cripps, Tristan Henderson, Patrick Proctor, Anna Shubina, and Jihwang Yeo. Jose Manuel Garc\'ia-Gim\'enez is acknowledged for his enthusiastic work on the FCParser. 

Some of the Dartmouth effort was funded through support from ACM SIGMOBILE and by an early grant from the US National Science Foundation under award number 0454062.
This work was also supported by the Agencia Estatal de Investigación in Spain, MCIN/AEI/ 10.13039/501100011033, grant No PID2020-113462RB-I00, and the European Union’s Horizon 2020 research and innovation programme under the Marie Skłodowska-Curie grant agreement No 893146. Funding for open access charge: Universidad de Granada / CBUA.


\setcounter{table}{0}

\bibliographystyle{elsarticle-num}
\bibliography{wanom}

\begin{IEEEbiography}[{\includegraphics[width=2.5cm]{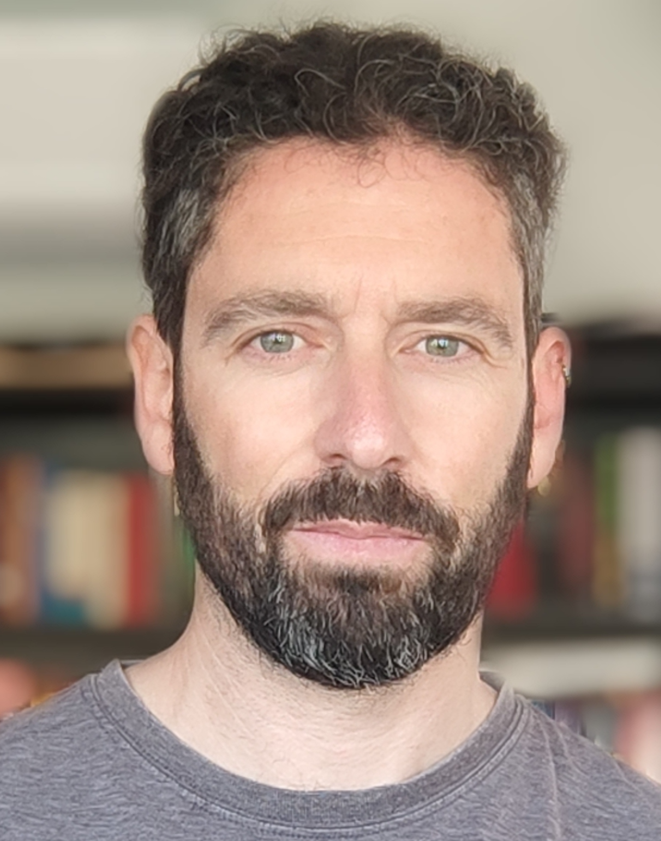}}]{José Camacho}
José Camacho is Full Professor in the Department of Signal Theory, Telematics and Communication and head of the Computational Data Science Laboratory (CoDaS Lab), at the University of Granada, Spain. He holds a degree in Computer Science from the University of Granada (2003) and a Ph.D. from the Technical University of Valencia (2007), both in Spain. He worked as a post-doctoral fellow at the University of Girona, granted by the Juan de la Cierva program, and was a Fulbright fellow in 2018 at  Dartmouth College, USA. His research interests include networkmetrics and intelligent communication systems, computational biology, knowledge discovery in Big Data and the development of new machine learning and statistical tools. 
\end{IEEEbiography}

\begin{IEEEbiography}[{\includegraphics[width=2.5cm]{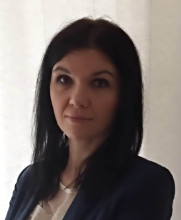}}]{Katarzyna Wasielewska} Katarzyna Wasielewska received MSc in computer science from the Faculty of Mathematics and Computer Science, Nicolaus Copernicus University in Torun (NCU), Poland, in 1999, and PhD in telecommunications from the Faculty of Telecommunication, Information Technology and Electrical Engineering, University of Science and Technology in Bydgoszcz (UTP), Poland, in 2014. She is Assistant Professor at the Institute of Applied Informatics, State University of Applied Sciences in Elblag, Poland. Currently, she is a Postdoctoral Researcher at the Department of Signal Theory, Telematics and Communication and the CoDaS Lab, University of Granada, Spain, granted by EU Marie Skodowska-Curie Actions Individual Fellowships program. Her research interests include computer communications, network traffic analysis, network security, multivariate analysis and machine learning. She worked 10 years as an ISP network administrator, and she is an active IEEE volunteer.
\end{IEEEbiography}

\begin{IEEEbiography}[{\includegraphics[width=2.5cm]{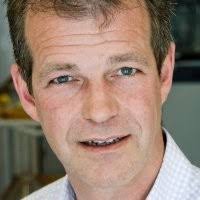}}]{Rasmus Bro}
Rasmus Bro (born 1965) studied mathematics and analytical chemistry at the Technical University of Denmark and received his M.Sc. in 1994. In 1998 he obtained his Ph.D. in multiway analysis from the University of Amsterdam, The Netherlands. Since 1994 he has been employed at the Department of Food Science, at the University of Copenhagen, and in 2002 he was appointed full professor of chemometrics. He has had several stays abroad at research institutions in The Netherlands, Norway, France, and United States. Current research interests include chemometrics, multivariate calibration, multiway analysis, exploratory analysis, blind source separation, curve resolution, MATLAB programming. 
\end{IEEEbiography}


\begin{IEEEbiography}[{\includegraphics[width=2.5cm]{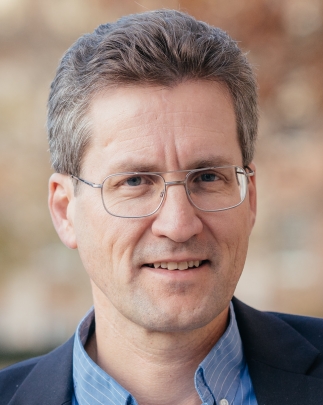}}]{David Kotz}
David Kotz is the Provost, and the Pat and John Rosenwald Professor in the Department of Computer Science, at Dartmouth College. He previously served as Associate Dean of the Faculty for the Sciences, as a Core Director at the Center for Technology and Behavioral Health, and as the Executive Director of the Institute for Security Technology Studies. His current research involves security and privacy in smart homes, and wireless networks. He has published over 250 refereed papers, obtained \$89m in grant funding, and mentored over 100 research students and postdocs. He is an ACM Fellow, an IEEE Fellow, a 2008 Fulbright Fellow to India, a 2019 Visiting Professor at ETH Zürich, and an elected member of Phi Beta Kappa. He received his AB in Computer Science and Physics from Dartmouth in 1986, and his PhD in Computer Science from Duke University in 1991.
\end{IEEEbiography}

\end{document}